\documentclass{IEEEtran}
\usepackage{bbm}
\usepackage{amssymb,amsmath,amsthm}
\usepackage{graphicx,epstopdf,psfrag,url,cite,xcolor,multirow,float}
\usepackage{caption-patch}
\usepackage[small,bf]{caption}
\usepackage[font=footnotesize]{subfig}

\usepackage{algorithm}
\usepackage{algorithmic}

\usepackage[super]{nth}
\usepackage{array}
\newcolumntype{C}[1]{>{\centering\let\newline\\\arraybackslash\hspace{0pt}}m{#1}}

\newtheoremstyle{mythm}
{\topsep}   
{\topsep}   
{\itshape}      
{0pt}       
{\bfseries} 
{:}         
{5pt plus 1pt minus 1pt}    
{\thmname{#1}\thmnumber{ #2}\thmnote{ (#3)}}
\theoremstyle{mythm}

\renewcommand{\algorithmicrequire}{\textbf{Input:}}
\renewcommand{\algorithmicensure}{\textbf{Output:}}

\allowdisplaybreaks[4]

\newcommand{\bm}[1]{\boldsymbol{#1}}

\IEEEoverridecommandlockouts

\begin{document}

\title{Bayesian Optimization Enhanced Deep Reinforcement Learning for Trajectory Planning and Network {Formation} in Multi-UAV Networks}

\author{Shimin Gong, Meng Wang, Bo Gu, Wenjie Zhang, Dinh Thai Hoang, and Dusit Niyato\\

\thanks{Meng Wang, Shimin Gong, and Bo Gu are with the School of Intelligent Systems Engineering, Shenzhen Campus of Sun Yat-sen University, China (email: wangm329@mail2.sysu.edu.cn, \{gongshm5, gubo\}@mail.sysu.edu.cn). Wenjie Zhang is with the School of Computer Sciences, Minnan Normal University, China (e-mail: zhan0300@ntu.edu.sg). Dinh Thai Hoang is with the School of Electrical and Data Engineering, University of Technology Sydney, Australia (email: hoang.dinh@uts.edu.au). Dusit Niyato is with School of Computer Science and Engineering, Nanyang Technological University, Singapore (email: dniyato@ntu.edu.sg).}
}

\maketitle
\thispagestyle{empty}
\begin{abstract}
In this paper, we employ multiple UAVs coordinated by a base station (BS) to help the ground users (GUs) to offload their sensing data. Different UAVs {can adapt} their trajectories and network formation to expedite data transmissions via multi-hop relaying. The trajectory planning aims to collect all {GUs' data, while the UAVs'} network formation optimizes the multi-hop UAV network topology to minimize the energy consumption and transmission delay. {The joint network formation and trajectory optimization is solved by} a two-step iterative approach. Firstly, we devise the adaptive network formation scheme by using a heuristic algorithm to balance the UAVs' energy consumption and data queue size. Then, with the fixed network formation, the UAVs' trajectories are further optimized by using multi-agent deep reinforcement learning without knowing the GUs' traffic demands and spatial distribution. To improve the learning efficiency, we further employ Bayesian optimization to estimate the UAVs' flying decisions based on historical trajectory points. This helps avoid inefficient action explorations and improves the convergence rate in the model training. The simulation results reveal close spatial-temporal couplings between the UAVs' trajectory planning and network formation. Compared with several baselines, our solution can better exploit the UAVs' cooperation in data offloading, thus improving energy efficiency and delay performance.
\end{abstract}


\section{Introduction}
Recently, unmanned aerial vehicles (UAVs) have been used in various {wireless networks} to provide channel access for wireless devices on the ground, constituting a significant part of the future Internet of Things (IoT), e.g.,~\cite{gupta2015survey} and~\cite{hayat2016survey}. Due to hardware constraints, a large portion of the low-cost IoT sensor devices may be difficult to meet the quality of service (QoS) requirements in data transmissions due to limited energy supply, remote deployment locations, and non-line-of-sight (NLoS) channel conditions. Such difficulties can be potentially resolved by deploying UAVs to assist wireless communications from the ground users (GUs) to remote base stations (BSs), e.g.,~\cite{ding2018amateur} and~\cite{zhao2019uav}. The UAVs' capability of fast deployment, mobility, and flexibility make it possible for UAV-assisted wireless networks to serve diverse IoT users in a large service area that is beyond the BS's direct coverage. It can not only extend the communication range but also improve the network capacity of dense IoT networks.

To explore the performance gain in UAV-assisted wireless networks, the UAVs' mobility should be jointly optimized with the {transceivers' control} strategies. In literature, the mobility of UAVs' has been exploited to improve the data rate, energy efficiency, and age of information (AoI) in wireless networks, e.g.,~\cite{yang2018optimal,mozaffari2017mobile,age2020Tran,abd2018average}. Thanks to a much better {GU-UAV (G2U)} direct channel condition, the UAVs can firstly fly to a point of interest, receive the information from the GUs, and then help forward the information to the remote BS. By planning the UAVs' flying trajectories, all GUs are expected to improve their data rate and reduce the transmission latency. As such, we can improve the overall network throughput and provide services to more GUs~\cite{yang2018optimal}. The UAVs can be also used to improve the energy efficiency of wireless networks by jointly optimizing the GUs' {transmit} power and the UAVs' trajectories~\cite{mozaffari2017mobile}. Instead of direct communications to the remote BS, the GUs can upload information to nearby UAVs with a higher data rate {and} a lower transmit power. Besides the improvement on network capacity and energy efficiency, UAVs can be employed to assist the GUs' information sensing and improve the AoI by planning the UAVs' trajectories to collect all GUs' sensing data timely. The authors in~\cite{abd2018average} revealed that AoI can be efficiently decreased by jointly optimizing the UAVs' trajectories, energy consumption, and the service time.

{However, the UAVs' trajectory optimization} is a high-dimensional control problem especially in a multi-UAV-assisted wireless network with complicated spatial-temporal couplings. {It can be} further complicated by the UAVs' energy constraints, the physical restrictions in flight control, and the GUs' diverse service requirements. For example, multiple UAVs can be dispatched to cover a large service area. {This} may consume more energy for distant UAVs to report their sensing data back to the BS. On the other hand, the UAVs can also operate in a small swarm to increase the {GUs' channel access probabilities}. However, this may decrease the energy efficiency and complicate the UAVs' transmission scheduling due to their mutual interference. Each UAV's trajectory planning not only affects its own energy- or information-efficiency, but also affects the other UAVs' {current and future trajectories}. Such spatial-temporal couplings among different UAVs make it more challenging to optimize the UAVs' task cooperation and trajectories to serve a large group of GUs in the same wireless network. The cooperation among UAVs can be realized by allowing the UAV-to-UAV (U2U) direct communications, e.g.,~\cite{zhang2019cellular} and~\cite{he2021multi}, leading to a multi-hop network of UAVs, namely the UAVs' network formulation. The UAV can first collect and cache the GUs' data and then forward the data to a nearby UAV when they meet each other on their trajectories. It is also possible for the UAVs to dynamically switch between different U2U links and thus adapt the network formation according to the UAVs' channel conditions, energy statuses, and data queue sizes.

The UAVs' joint trajectory optimization and adaptive network formation have not been well studied in the literature. Given the UAV's network formation, the trajectory optimization can be discretized in different time slots and then formulated as the mobility control problem in each time slot~\cite{zhang2019dual}. {The optimization methods often require} complete information about the GUs' distribution and traffic demands. This makes it more difficult for practical deployment in a dynamic network environment. The model-free deep reinforcement learning (DRL) method is also a promising technique to optimize the UAVs' trajectories in wireless networks, e.g.,~\cite{drl_survey} and~\cite{RL-resource-allocation}. {It can} adapt the UAVs' trajectories by interacting with the network environment. A typical implementation of the DRL algorithm relies on the centralized control at the BS, which collects information from all UAVs and jointly adapts their trajectories in each time step to maximize the overall network performance. This may require excessive {communications} and training overheads as the number of UAVs increases. On the other hand, the UAVs can be regarded as independent decision-making agents, which can adapt their trajectories based on local observations of the network environment. However, such a multi-agent decentralized implementation still requires a centralized training which can be costly in terms of the communication overhead. The above challenges motivate our work in this paper to design a more efficient algorithm to jointly optimize the UAVs' trajectories and network formation strategy that fulfill the GUs' traffic demands in a dynamic wireless network.

In this paper, we focus on a multi-UAV-assisted wireless network and explore the performance gain via the UAVs' cooperation. We aim to minimize the time delay and energy consumption for data collection by jointly optimizing the UAVs' trajectories and network formation, which are typically viewed as two different design problems and tackled separately~\cite{zhang2019cellular}. 
Our analysis reveals that the UAVs' trajectories and network formation are closely dependent on {each other}. We expect that the UAVs' trajectories can be very different according to the GUs' spatial distribution and traffic demands. When two UAVs fly far apart, their U2U channel becomes deteriorated and even disconnected. This implies that the UAVs' network formation {should be} adaptive to the change of the UAVs' trajectories. To this end, we propose a two-step algorithm to iterate between the UAVs' trajectory optimization and adaptive network formation. The basic idea of the adaptive network formation is to evaluate the UAVs' local resource consumption and ensure {load balance} among different UAVs. Once the UAVs' trajectories in current time slot become instable {and exaggerate the unbalanced resource consumption at the UAVs}, the BS will initialize the adaptive network formation to optimize the UAVs' network topology or channel allocation. The change of network formation further urges each UAV to update its trajectory, which is addressed by using the multi-agent deep deterministic policy gradient (MADDPG) algorithm~\cite{ekram-madrl}. Moreover, to improve the learning efficiency, we propose an action estimation mechanism by using Bayesian optimization to estimate more rewarding action for each UAV. Based on the UAVs' past trajectories, the action estimation can avoid ineffective action exploration and potentially improve the learning performance. Our simulation results verify that the joint trajectory planning and network formation algorithm can significantly reduce the transmission delay compared with several baseline strategies. The adaptive network formation is effective to improve the network performance by exploiting the UAVs' cooperation. The Bayesian optimization enhanced MADDPG algorithm also shows improved learning efficiency, stability, and reward performance compared to the conventional MADDPG algorithm.

Specifically, our main contributions in this paper are summarized as follows:
\begin{itemize}
\item Joint trajectory planning and network formation: Multiple UAVs are employed to collect data from the GUs and help forward the sensing data to the remote BS. The UAVs' adaptive network formation aims to balance {the traffic load and} resource consumption among different UAVs by allowing U2U multi-hop relaying communications. We propose to jointly optimize the UAVs' trajectories and network formation to exploit the UAVs' cooperation gain.
\item Energy- and delay-aware adaptive network formation: We propose the two-step solution to update the UAVs' trajectories and network formation iteratively. Once the UAVs' trajectories lead to unbalanced resource consumption, the network formation will be performed to restore the resource balance among UAVs. The trajectory planning is based on the MADDPG algorithm, while the { network formation} is updated by a heuristic algorithm driven by the UAVs' status information including the energy supply, channel conditions, and data queue size.
\item Bayesian optimization enhanced MADDPG: We improve the learning efficiency of {the conventional MADDPG} by using Bayesian optimization to estimate the UAVs' flying locations based on the past trajectories. This can guide the UAVs' trajectory learning towards a more rewarding policy. The DRL agent's action exploration also provides more sample information for Bayesian optimization to make accurate action estimation. Extensive simulation results verify that the proposed learning framework significantly improves the learning performance compared to the conventional MADDPG algorithm.
\end{itemize}

Some preliminary results of this work have been {appeared in a short conference paper}~\cite{9780862}. This paper further extends the study in~\cite{9780862} by proposing Bayesian optimization method to improve the learning performance of the MADDPG algorithm for the UAVs' trajectory planning. Extensive simulation results are also provided to verify that the adaptive network formation along with the UAVs' trajectories can balance the UAVs' resource consumption and minimize the latency in data collection. The rest of this paper is organized as follows. Section~\ref{sec-relate-work} provides a discussion on related works. We present the system model in Section~\ref{sec-model} and the solution framework in Section~\ref{sec-problem}. In Section~\ref{sec-bayesian}, we further employ Bayesian optimization for action estimation to improve the learning efficiency. Finally, numerical results and conclusions are presented in Sections~\ref{sec-simulation} and~\ref{sec-conclusion}, respectively.

\section{Related Works}\label{sec-relate-work}

\subsection{Multi-UAV Cooperative Networks}
{One major problem} in UAV-assisted wireless networks is to optimize the UAVs' trajectories to maximize the network performance, such as service coverage, overall network throughput, and transmission delay. A large portion of the existing works focused on a \emph{non-cooperative} case by considering the direct links between UAVs and the remote BSs, aiming to optimize individual UAVs' trajectories and transmission control strategies. For example, the authors in~\cite{zhan2017energy} employed the UAV to collect data from GUs with fixed locations and aimed to minimize the GUs' energy conconflictsumption. The authors in~\cite{alzenad20173} studied the UAV's placement strategy to maximize the number of GUs under its coverage. {Without considering the U2U connections, the authors in~\cite{Coalition-Forma-UAV} proposed the coalition formation game to study the UAVs' task assignment problem in a large service area. Each UAV will be allocated a dedicated service area to avoid resource conflict. Such} non-cooperative strategies can be easy to implement, but the direct {UAV-BS (U2B)} links may limit the service range of multi-UAV networks. \emph{Task cooperation} among UAVs can be realized by using multi-hop U2U relay communications to enhance the data collection for distant GUs~\cite{frew2008airborne}. The authors in~\cite{challita2017network} proposed a backhaul scheme that uses UAVs to forward sensing information to the BS. A network formation game is formulated to construct the UAV backhaul by multi-hop communications among different UAVs. The authors in~\cite{zhang2019cellular} further proposed a sub-channel allocation strategy to improve the throughput performance of the network formation, by a joint optimization of the UAVs' channel assignment and flight control. {The authors in~\cite{Resource-Opt} considered a multi-hop UAV-assisted relay network to assist a set of transceivers pairs on the ground to communicate with each other. The minimum transmission rate can be maximized by optimizing the UAVs' deployment locations, the transmit power, and bandwidth allocation.} Similarly, the authors in~\cite{he2021multi} focused on the multi-hop UAV-assisted task offloading system. The UAVs' resource allocation and deployment strategy are jointly optimized to maximize the on-the-fly computation capability. Considering {the} UAVs' limited processing capabilities, a distance-based UAV cooperation scheme was proposed in~\cite{xia2022intelligent} by allowing each UAV to seek assistance from the closest neighboring UAV. A buffer-based routing strategy was proposed in~\cite{buffer-size} for a multi-hop MEC network. A similar routing problem was studied in a UAV-assisted multi-hop MEC network~\cite{huawei}, by optimizing the UAVs' computation offloading and multi-hop routing strategies. {To save the UAVs' energy consumption in data aggregation, the authors in~\cite{Data-Aggregation} proposed different routing methods for the UAVs to transmit data to the receiver. Each UAV can either rely on hop-by-hop routing algorithm or form a coalitional structure by the coalition formation game algorithm, based on the UAVs' data demands, locations, and network topology.} The multi-UAV network is expected to handle more complex sensing tasks and meet heterogeneous requirements by allowing the adaptive network formation according to their network statuses, including the network topology, channel condition, buffer size, energy status, etc. This motivates our study in this work.

\subsection{DRL for UAV-assisted Wireless Networks}
By interacting with the environment, the DRL agent can adjust the UAVs' trajectories and data offloading strategies based on time-varying workload demands and channel conditions. Considering multiple GUs, the authors in~\cite{chu2021fast} employed the Q-learning algorithm to adapt the UAV's speed control, based on its energy status and position. The authors in~\cite{9513250} proposed the actor-critic method to optimize the data collection from GUs by using multiple UAVs. The authors firstly employed the k-mean clustering to aggregate different GUs and then optimize the UAVs' trajectories by the actor-critic method. The actor-critic DDPG was employed in~\cite{bouhamed2020uav} to jointly optimize the UAVs' trajectories and transmission scheduling strategy. Besides throughput maximization, the UAVs' fast deployment can also help reduce the transmission delay or age of information (AoI). The authors in~\cite{aoi-wpt-2022} studied the AoI minimization in a UAV-assisted wireless network with RF power transfer. The DQN approach was proposed to adapt the UAV's trajectory, the GUs' scheduling and energy harvesting policies. The authors in~\cite{aoi-energy-uav} studied the AoI-energy-aware data collection in UAV-assisted wireless networks. The TD3 algorithm is proposed to minimize the weighted sum of average AoI, the UAV's propulsion energy, and the GUs' transmission energy, by jointly optimizing the UAV's flying speed, hovering locations, and bandwidth allocation for data collection.

\subsection{Multi-agent DRL for Trajectory Optimization}
The multi-agent DRL (MADRL) framework has also been proposed in literature to optimize the UAVs' trajectories for data collection. The authors in~\cite{cui2019multi} developed the multi-agent DQN method, which is a simple extension of the DQN method to multi-UAV scenario. The authors in~\cite{Dynamic-Wireless} viewed each UAV as an independent DQN agent that makes decision based on its local observation to maximize the real-time downlink capacity while covering all GUs. The authors in~\cite{qie2019joint} proposed the MADDPG method to optimize the UAVs' target assignment and trajectory planning. The MADDPG method was also employed in~\cite{Enabled-Secure-Commu} to ensure secure communication in UAV-assisted wireless networks. The authors in~\cite{uavs-aoi} employed multiple UAVs to help collect sensing information from a set of GUs. The actor-critic DRL approach was proposed to minimize the AoI by adapting each UAV's sensing and {flying decisions}. A similar problem was studied in~\cite{maddpg-aoi-uav} by using the MADDPG algorithm to optimize the UAVs' trajectories and transmission strategy. The multi-agent actor-critic method was proposed in~\cite{mec-aoi} to adapt the UAVs' actions including the {mobility control} strategy, the computing resource allocation, and offloading scheduling decisions. The authors in~\cite{zhang2020hierarchical} studied the multi-UAV-assisted data collection where the UAVs can help activate the GUs and then collect the data via backscatter communications. Each UAV can also fly back to a charging station to charge itself when its energy becomes low. A multi-agent deep option learning (MADOL) algorithm was proposed to minimize the UAVs' total flight time by learning the UAV-GUs association strategy.

\section{System Model}\label{sec-model}

\begin{figure}[t]
\centering
\subfloat[Multi-UAV-assisted data offloading to the remote BS.]{\includegraphics[width=1\linewidth]{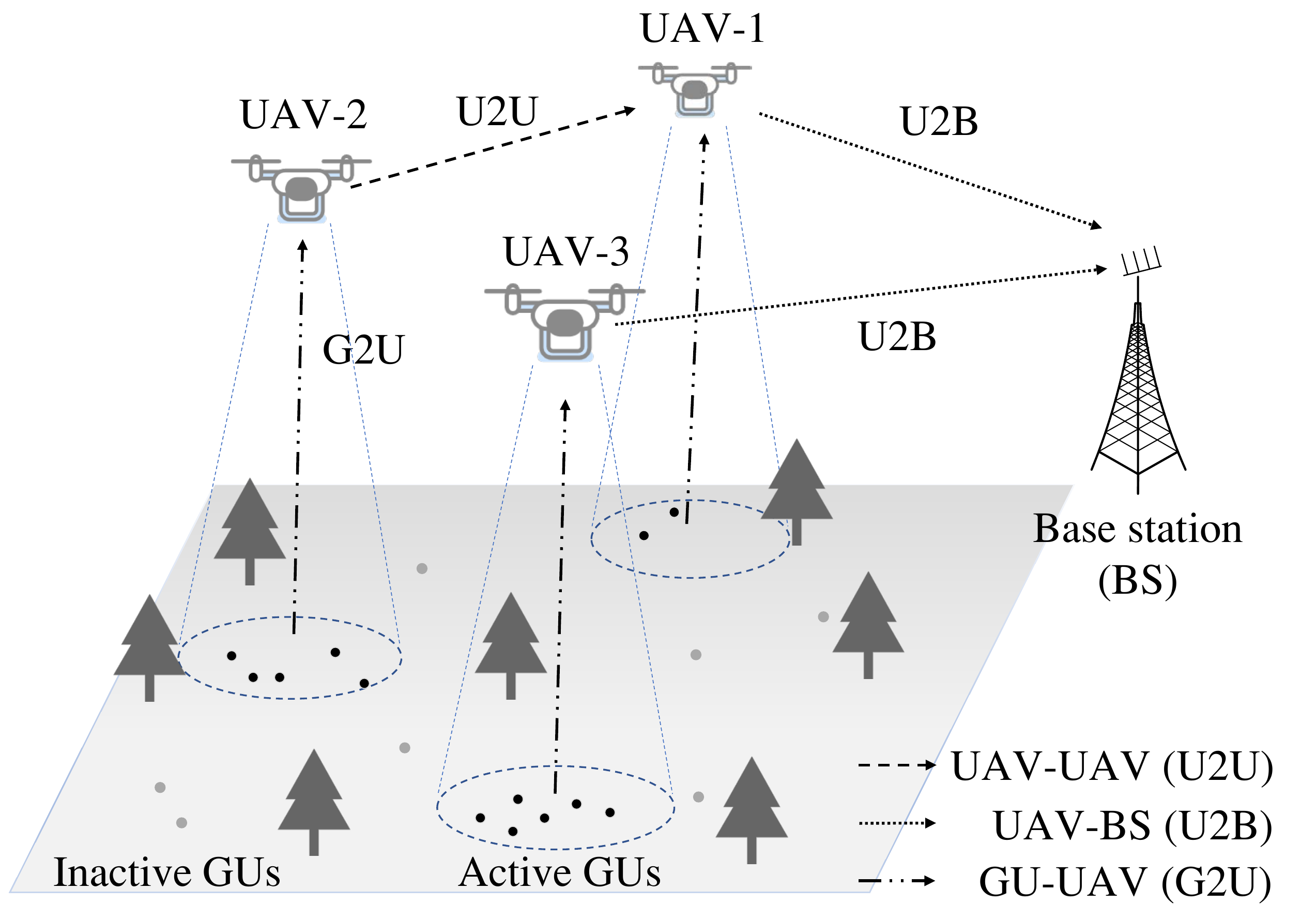}}\\
\subfloat[Time-slotted Fly-Sense-Offload protocol.]{\includegraphics[width=1\linewidth]{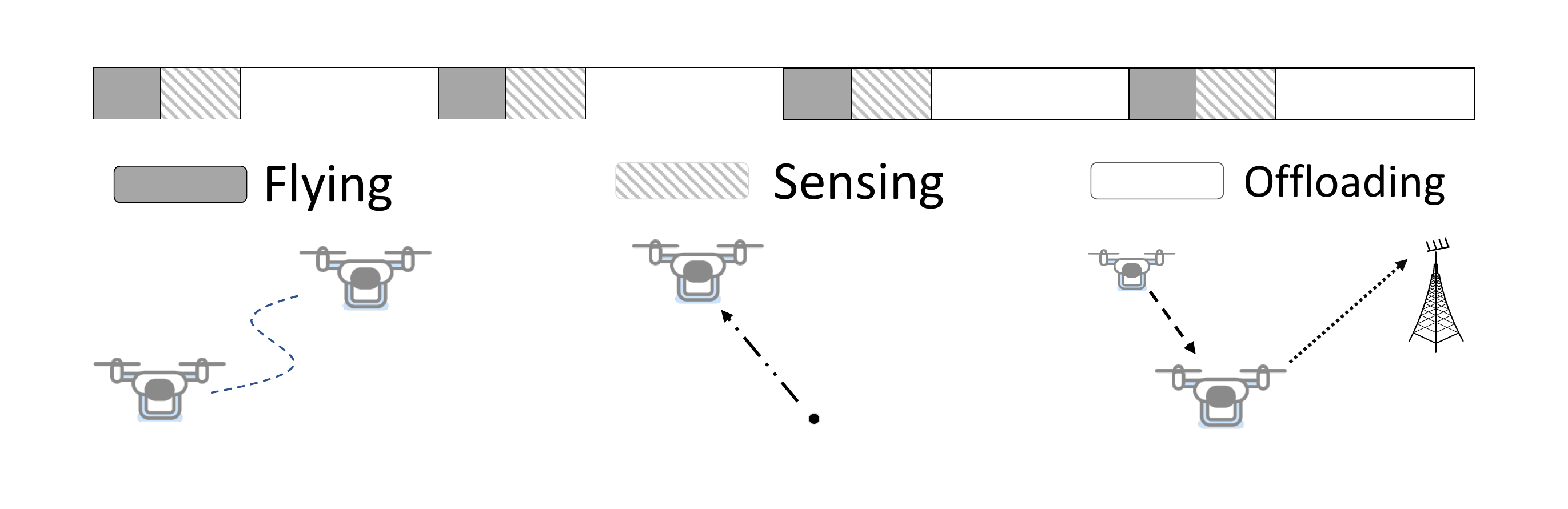}}
\caption{System model}
\label{fig-model}
\end{figure}

As illustrated in Fig.~\ref{fig-model}, we consider a multi-UAV-assisted wireless network with one BS and multiple UAVs, denoted by the set $\mathcal{N}= \{1, 2,\ldots, N\}$. There are a set of sensors or IoT user devices, denoted as~$\mathcal{M} = \{ 1, 2, \ldots, M\}$, spatially distributed on the ground and may beyond the direct communication range with the remote BS. Multiple {UAVs can} fly around and collect sensing data from the GUs. The collected data is firstly buffered at the UAVs and {then can be} forwarded to the BS when the UAVs' channel conditions become more preferable. Each UAV has a limited buffer size $D_{\max}$. Buffer outage happens when the collected sensing data exceeds the UAV's buffer size. Depending on the UAV's channel conditions and energy status, each UAV can either connect to the BS directly or relay its {information to} the BS through the other UAVs. The {direct U2U} communications allow the UAVs to form a multi-hop network topology, namely the network formation, and thus potentially decrease the overall transmission delay and energy consumption by multi-hop relaying communications. 


\subsection{Time-slotted Fly-Sense-Offload Protocol}

We consider a time-slotted frame structure as shown in Fig.~\ref{fig-model}(b). In each time slot $t \in \mathcal{T}\triangleq\{1,2,\ldots\}$, each UAV can fly to a different location, collect the sensing data from the GUs, and then forward the sensing data to the BS or another UAV. Each time slot has unit length and can be further divided into three sub-slots, i.e.,~the UAV flying sub-slot $t_{f}$, data sensing $t_{s}$, and offloading sub-slots $t_{o}$. In the flying sub-slot $t_{f}$, the UAV will fly from one location to another with a controllable speed $v_i$. In the sensing sub-slot $t_{s}$, the UAV will collect the sensing data from the GU with the strongest sensing signal within the UAV's service coverage. Each GU $m\in\mathcal{M}$ has a fixed amount of data transmission demand $W_m$ and it will stop transmission when the remaining data size becomes zero or the UAVs' buffer spaces become full. In the offloading sub-slot $t_{o}$, the UAV can offload a part of its data either to the BS directly or to a nearby UAV with more preferable channel conditions and energy supply. {During data offloading, each UAV can also update its own status information to the BS, including its location, buffer size, energy status, and channel conditions. The collection of all UAVs' status information will help the BS to adapt the UAVs' trajectories and transmission control strategies.} For simplicity, we assume that the sensing sub-slot $t_{s}$ is fixed, while the flying and offloading sub-slots can be jointly optimized to improve the transmission efficiency. The time-slotted Fly-Sense-Offload protocol for multi-UAV-assisted data offloading allows three types of communications links among GUs, UAVs, and the BS, detailed as follows:
\begin{enumerate}
\item {The G2U channels are used for the UAVs to collect} data from the GUs. We assume that the direct channels from the GUs to the BS are not available due to physical obstacles. All sensing data will be collected by {the UAVs} and re-routed to the BS. Each GU can successfully transmit its data to the UAV when the received signal-to-noise ratio (SNR) meets the minimum threshold.
\item {The U2B channels can be used by the UAVs to communicate directly  with the BS} if the UAVs are close to the BS. We assume that the U2B communications rely on a set of dedicated reporting channels which are shared among all UAVs. {The data rates of the U2B channels depend} on the UAVs' locations, transmit power, and channel conditions.
\item {The U2U channels can be used to connected nearby UAVs} in the case that some UAVs are far away from the BS. {The GUs' sensing data can be forwarded to the BS via multi-hop UAV relaying communications.}
\end{enumerate}
We also assume that each UAV has a single antenna and thus it can transmit data by using either the U2B or U2U channel. Moreover, the U2U and U2B channels are sharing the same spectrum resources. Considering limited channel resources, here, all UAVs share $K$ orthogonal sub-channels, denoted by the set $\mathcal{K}= \{1,2,\dots,K\}$. { Once the G2U link is established, the GU will upload its sensing data to the UAV in the sensing sub-slot $t_s$. After data collection, the UAVs will update their strategies in the flying sub-slot $t_o$. Given the UAVs' hovering positions, each UAV can broadcast a pilot beacon signal to the GUs under its signal coverage. The active GUs with traffic demands can reply to the UAV with equal transmit power. Then, the UAV can estimate the signal qualities of different GUs and select the GU with the highest signal strength for uplink data transmission. By this way, each UAV will gain the information about the GUs' traffic demands. The uplink data transmission in each sensing sub-slot can be also extended to multiple access scenarios. When multiple GUs are selected to upload their data in the same sensing sub-slot $t_s$, the UAV can employ the non-orthogonal multiple access (NOMA) technique or time division multiple access (TDMA) protocol to coordinate the GUs' uplink data transmissions, which is beyond our discussion in this paper.}

\subsection{U2U Links and Network Formation}


{The U2U connections allow the UAVs to form a multi-hop UAV network to forward all sensing data to the BS. We call such a multi-hop backhaul as the UAVs' network formation, which specifies the feasibility of U2U connections among UAVs.} Given the UAVs' network formation, each UAV can {optimize its trajectory and} then forward the buffered data via the U2U or U2B channels. However, as the {UAVs change hovering locations}, the UAVs' network formation may become obsolete due {to the} change of U2U channel conditions. Thus, the UAVs' network formation has to be jointly optimized with the UAVs' trajectories. Specifically, it should be adapted dynamically according to the UAVs' channel conditions, energy supply, buffer size, and locations. For example, when the UAVs are distant from the BS, the direct U2B channels may experience a low SNR and a larger transmission delay, which implies a longer hovering time and a higher energy consumption. In this case, the UAVs can change the network formation by using the U2U channels and connecting with each other in a multi-hop relay network. In another case, when the GUs' data traffic is unevenly distributed, some UAVs may collect a large amount of sensing data while the others have little sensing data. Such an unbalanced traffic load will cause congestion in data transmission and potentially increase the transmission delay. {As such}, the UAVs with heavy load can offload data to the other UAVs by using the high-speed U2U channels. 

For notational convenience, let UAV-$0$ denote the BS with the fixed location and let $ \tilde{\mathcal{N}} = \mathcal{N}\cup \{0\}$ denote the set of all UAVs including {the UAV-$0$}. We define the binary matrix $\Phi(t) = [\phi^k_{i,j}(t)]_{i, j \in \tilde{\mathcal{N}}, k \in \mathcal{K}}$ to denote the U2U sub-channel allocation strategy, i.e., $\phi_{i,j}^k(t) = 1$ means that the $k$-th sub-channel is used for {the U2U channel between the UAV-$i$ and UAV-$j$}. It is easy to see that $\phi_{0,j}^k(t) = 0$ and $\phi_{j,j}^k(t) = 0$ for all $j \in \tilde{\mathcal{N}}$. We require that each sub-channel can be used for either information transmission or reception. Thus, the UAVs' sub-channel allocation is subject to the following constraint:
\begin{equation}\label{equ-channel-alloc}
\begin{aligned}
\sum_{m\neq i, m\in{\mathcal{N}}} \phi_{m,i}^k(t) + \sum_{j\neq i, j\in\tilde{\mathcal{N}}} \phi_{i,j}^k(t) \leq 1 , \forall \,i\in\tilde{\mathcal{N}}, k\in\mathcal{K}.
\end{aligned}
\end{equation}
It is clear that the feasible set for $\Phi(t)$ specifies all possible network formation structures for the UAVs.

\subsection{Channel Models and Data Offloading Rates}
We assume that all UAVs fly at a fixed altitude $H$ to collect the sensing data from the GUs. Our problem formulation and solution can be easily extended to the case with a time-varying flying altitude. Each UAV-$i$'s trajectory can be defined as a set of location points over different time slots, i.e.,~$\mathcal{L}_i=[\boldsymbol{\ell}_i(t)]_{t\in\mathcal{T}}$, where the location $\boldsymbol{\ell}_i(t)$ in each time slot is specified by three-dimensional (3D) coordinate, i.e,~$\boldsymbol{\ell}_i(t) = (x_i(t), y_i(t), z_i(t)=H)$. The BS locates at the origin of coordinate and the height of antenna is given by $H_b$. Given that the UAV-$i$ moves in the direction $\boldsymbol{d}_i(t)$ with a limited speed $v_i(t)\leq v_{\max}$, the UAV-$i$'s location in the next time slot $t+1$ is given by $\boldsymbol{\ell}_i(t+1) = \boldsymbol{\ell}_i(t) + v_i(t)\boldsymbol{d}_i(t)$. The distance between UAV-$i$ and UAV-$j$ is given by $d_{i,j}(t) = ||\boldsymbol{\ell}_i(t) - \boldsymbol{\ell}_j(t) ||$.

Typically, it is line-of-sight (LoS) transmission between UAVs and the BS. When the UAV-$i$ forwards sensing data to the UAV-$j$ on the sub-channel $k\in\mathcal{K}$, the received signal power at the UAV-$j$ can be denoted as $p_{j,i}^k(t) = p_i^k\beta_{i,j}^u (d_{i,j}(t))^{-\alpha_u}$, where $p_i^k$ denotes the UAV-$i$'s transmit power on the $k$-th sub-channel and $\beta^u_{i,j}$ is a constant power gain induced by the transceivers' amplifier and antenna. The path loss $(d_{i,j}(t))^{-\alpha_u}$ depends on the distance between the transceivers and $\alpha_u$ denotes the path-loss constant. As all UAVs share the same set of channels, the interference may be incurred between different {transceivers}. In particular, if UAV-$m$ (for $m\neq i$) also transmits on the sub-channel $k$, the interference to UAV-$j$ is given as follows:
\begin{equation}\label{equ-interf-u2u}
I_{j,i}^k(t) =  \sum_{m\neq i} \sum_{n\neq j} \phi^k_{m,n}(t)p_m^k(d_{m,j}(t))^{-\alpha_u}.
\end{equation}
Hence, the offloading rate from the UAV-$i$ to the UAV-$j$ (for $i,j\in\tilde{\mathcal{N}}$) {over all} sub-channels is determined as follows:
\begin{equation}\label{equ-u2u-rate}
o_{i,j}(t) = \sum_{k\in\mathcal{K}} \phi^k_{i,j}(t)\log\left(1+ \frac{p_{j,i}^k(t)} {\delta_k^2 + I_{j,i}^k(t)} \right),
\end{equation}
where $\delta_k^2$ denotes the noise power on the $k$-th sub-channel.

Similarly, we can define the channel model from the GUs to the UAVs. We assume that each UAV only collects data from the GUs in its coverage with LoS channel conditions. Hence, we can employ a similar log-distance path loss model as that for the {U2U and U2B channels. Let} $q_m$ denote the GU-$m$'s transmit power and $d_{m,i}(t)$ denote the distance between the GU-$m$ and the UAV-$i$. The data rate from the GU-$m$ to the UAV-$i$ is $u_{i,m}(t) = \log\left(1+  { q_m\beta^s_{m,i} (d_{m,i}(t))^{-\alpha_s} } \right)$, where $\alpha_s$ is the path-loss constant and $\beta^s_{m,i}$ denotes the channel gain at a reference point normalized by the noise power at the receiver. Note that we omit the mutual interference among different GUs. This is reasonable as different UAVs will stay at different locations to collect the GUs' sensing data. The UAVs' spatial separation avoids the GUs' interference as they upload information to different UAVs.

\subsection{Data Queue Dynamics at GUs and UAVs}\label{sec:Data Queue Dynamics}

We aim to optimize the UAVs' trajectories and network formation to minimize the transmission delay and  overall energy consumption. Initially, the GU-$m$ has a fixed amount of sensing data $W_m$ that needs to be offloaded to the BS. Given the fixed sensing sub-slot $t_s$, the UAV-$i$ can collect a part of the sensing data from the GUs, and then forward the buffered data to a nearby UAV or to the BS in the offloading sub-slot $t_o$. Therefore, there is a dynamic update of the GUs' and the UAVs' data queues over time. For each GU-$m$, its data queue can be updated as follows:
\begin{equation}\label{equ-iot-buffer}
W_m(t+1) = \left[W_m(t) - \sum_{i\in\mathcal{N}}x_{i,m}(t)s_{i,m}(t) \right]^+,
\end{equation}
where $[X]^+ = \max\{0, X\}$ and we define $s_{i,m}(t)$ as the amount of the GU-$m$'s sensing data collected by the UAV-$i$. The binary variable $x_{i,m}(t)\in\{0,1\}$ denotes the association between the UAV-$i$ and the GU-$m$ in the $t$-th time slot, i.e.,~{the GU-$m$'s sensing} data $s_{i,m}(t)$ will be collected by the UAV-$i$ if $x_{i,m}(t)=1$. Normally, we require that each GU connects to at most one UAV in each time slot~\cite{zhang2019cellular}, i.e.,~$\sum_{i\in\mathcal{N}}x_{i,m}(t)\leq 1$, due to the GU's limited transmission capability.

Let $\mathcal{M}_i$ denote the set of GUs in the UAV-$i$'s coverage. Then, for each UAV-$i$, the size of sensing data collected from the GUs can be denoted as follows:
\begin{equation}\label{equ-s-uav}
s_{i}(t) = \sum_{m\in\mathcal{M}_i} x_{i,m}(t) s_{i,m}(t).
\end{equation}
Besides the new sensing data $s_{i}(t)$, the UAV-$i$ may receive data from the other UAVs. Let $\mathcal{I}_i(t) \triangleq s_i(t) + \sum_{j\neq i, j\in\mathcal{N}} o_{j,i}(t)$ denote the incoming data to the UAV-$i$'s buffer space. In the offloading sub-slot $t_o$, the UAV-$i$ will forward its data either to a nearby UAV or to the BS. We define $O_{i}(t)$ as follows to denote the out-going data from the UAV-$i$:
\begin{equation}\label{equ-outgoing}
O_{i}(t) \triangleq o_{i,0}(t) + \sum_{j\neq i, j\in\mathcal{N}} o_{i,j}(t),
\end{equation}
where the offloading rate $o_{i,j}(t)$ is defined in~\eqref{equ-u2u-rate}. The first term $o_{i,0}(t)$ in~\eqref{equ-outgoing} is the data sent to the BS while the second term $\sum_{j\neq i, j\in\mathcal{N}} o_{i,j}(t)$ denotes the data forwarded to the other UAVs. Hence, the UAV-$i$'s data queue dynamics can be represented as follows:
\begin{equation}\label{equ-data-buffer}
D_i(t+1) =\min\left\{\left[D_{i}(t) - O_{i}(t) \right]^+ + \mathcal{I}_i(t) ,D_{\max}\right\}.
\end{equation}
In this paper, we consider a simple data collection {strategy, i.e., each} UAV is associated with the GU with the best signal quality under its coverage. Hence, the set $\mathcal{M}_i$ always contains one GU with the strongest signal strength to the UAV-$i$, say the GU-$m$, which can be easily determined given the UAV-$i$'s location. Given the sensing time $t_s(t)$, the size of sensing data can be evaluated as follows:
\begin{equation}\label{equ-g2u}
s_{i,m}(t)  = t_{s}(t)\log\left(1+  { q_m\beta^s_{m,i} (d_{m,i}(t))^{-\alpha_s} } \right).
\end{equation}

\section{Problem Formulation and Learning-based Solution}\label{sec-problem}

After collecting the sensing data, each UAV-$i$ will forward the buffered data to a nearby UAV or to the BS during its data offloading sub-slot $t_{i,o}$. Then, it will fly to the next point in the flying sub-slot $t_{i,f}$. Each UAV-$i$'s energy consumption mainly depends on its flying speed $v_i(t)$, hovering time $t_{i,o}$, and the flying time $t_{i,f}$ in the air. We can characterize the UAV-$i$'s operational energy consumption $e_i(t)$ in the $t$-th time slot by using the well-know energy model in~\cite{zeng2017energy}. Besides, let $p_i(t) = \sum_{j\in\mathcal{N}}\sum_{k\in\mathcal{K}}t_{i,o}p_{i,j}^k \phi_{i,j}^k(t)$ denote the UAV-$i$'s energy consumption on data offloading. Hence, the UAV-$i$'s overall energy consumption in each time slot is given by $\hat{e}_{i}(t) = e_i(t) + p_i(t)$. Till this point, we can formulate the energy minimization problem as follows:
\begin{subequations}\label{prob-mini}
\begin{align}
\min_{X, \Phi,\mathcal{L}_i, T} ~&~ \sum_{i=1}^N \sum_{t=1}^T\left( e_i(t) + \sum_{j\in\mathcal{N}}\sum_{k\in\mathcal{K}}t_{i,o}p_{i,j}^k \phi_{i,j}^k(t) \right) \label{obj-energy}\\
s.t.
~&~ \eqref{equ-channel-alloc}-\eqref{equ-g2u},\label{con-model}\\
~&~ D_i(t) \leq D_{\max} \text{ and } D_i(T) = 0 ,\label{con-uav-buffer}\\
~&~ W_m(0) = D_m \text{ and } W_m(T) = 0, \label{con-iot-buffer} \\
~&~ ||\boldsymbol{\ell}_i(t+1) - \boldsymbol{\ell}_i(t) || \leq v_{\max}(t) t_{i,f},  \label{con-uav-fly}\\
~&~ ||\boldsymbol{\ell}_i(t) - \boldsymbol{\ell}_j(t) || \geq d_{\min},  \label{con-uav-safe}\\
~&~ \phi^k_{i,j}(t) \in \{0,1\}  \text{ and }  x_{i,m}(t)  \in \{0,1\}, \label{con-formation} \\
~&~ \forall\, i, j \in\mathcal{\mathcal{N}}, \forall\, t\in\mathcal{T} \text{ and } \forall\, m\in\mathcal{M}, k\in\mathcal{K}.
\label{con-all-ind}
\end{align}
\end{subequations}

We aim to optimize the network formation $\Phi(t)$ and the binary matrix $X(t)=[x_{i,m}(t)]_{i\in\mathcal{N}, m\in\mathcal{M}}$ that specifies the G2U association strategy in each time slot $t\in\mathcal{T}$. All these matrix variables should be jointly optimized with the UAVs' trajectories $\mathcal{L}_i$ for $i\in\mathcal{N}$. We also optimize the total number of time slots $T$ that is required to complete all GUs' data offloading to the remote BS. For simplicity, we can consider a fixed data collection strategy in this paper as detailed in Section~\ref{sec-model}-D, i.e.,~each UAV-$i$ only collects the sensing data from the GU with the highest signal strength. As such, the G2U association matrix $X(t)$ can be known given the UAVs' locations in each time slot. The constraints in~\eqref{equ-channel-alloc}-\eqref{equ-g2u} specify the sub-channel allocation strategy and the buffer dynamics in both UAVs and GUs. The constraints in~\eqref{con-uav-buffer}-\eqref{con-iot-buffer} ensure that all GUs' sensing data can be successfully offloaded to the BS after $T$ time slots. The inequalities in~\eqref{con-uav-fly} and~\eqref{con-uav-safe} restrict the UAVs' trajectories in different time slots. Practically, the UAVs' transmit power in the objective~\eqref{obj-energy} is much less than the power consumption for the UAV's hovering and flying, and thus can be omitted in the optimization problem.

Problem~\eqref{prob-mini} is a mixed integer problem and difficult to solve efficiently due to spatial and temporal couplings between the UAVs' { network formation} and trajectory planning. The optimization of the time span $T$ further makes it inflexible for problem reformulation. Given a fixed number of time slots, the UAVs and GUs may have remaining data in their buffer spaces, i.e.,~the constraints in~\eqref{con-uav-buffer}-\eqref{con-iot-buffer} may not hold at the end of $T$ time slots. As such, we revise the objective in~\eqref{obj-energy} to take account the remaining data in buffers as penalty terms and reformulate~\eqref{prob-mini} into a joint optimization problem as follows:
\begin{subequations}\label{prob-mini-new}
\begin{align}
\min_{\Phi,\mathcal{L}_i} ~&~ \sum_{t=1}^T\left(\sum_{i=1}^N \Big( \hat{e}_i(t) + \lambda_i D_i(t) \Big )  + \sum_{m=1}^M W_m(t)\right)\\
s.t.
~&~ \eqref{con-model} \text{ and } \eqref{con-uav-fly}-\eqref{con-all-ind},
\end{align}
\end{subequations}
where $\lambda_i$ is constant parameter to trade off between the UAVs' energy consumption and data queue sizes. {Here we omit the UAVs' transmit power consumption in~\eqref{obj-energy}, which is much less than the power consumption for hovering and flying}. Given a fixed number of $T$ time slots, we aim to jointly minimize the overall energy consumption and the queue sizes of both UAVs and GUs. In the sequel, we devise an approximate solution to problem~\eqref{prob-mini-new} by decomposing it into the UAVs' network formation and the trajectory planning sub-problems. 

\subsection{Adaptive Network Formation}
In the first sub-problem, {we adapt} the UAVs' network formation $\Phi(t)$ on demand given the UAVs' trajectories $\mathcal{L}_i$. Hence, we can further simplify problem~\eqref{prob-mini-new} as follows:
\begin{subequations}\label{prob-network}
\begin{align}
\min_{\Phi} ~&~ \sum_{t=1}^T\left(\sum_{i=1}^N \Big( \hat{e}_i(t) + \lambda_i D_i(t) \Big )  + \sum_{m=1}^M W_m(t)\right) \label{obj-network}\\
s.t.
~&~ \eqref{con-model} \text{ and } \eqref{con-formation}-\eqref{con-all-ind}.
\end{align}
\end{subequations}
Problem~\eqref{prob-network} becomes a nonlinear integer program. Though it can be solved by the existing branch-and-bound method, it has a very high computational complexity due to the dynamic evolution of the UAVs' and the GUs' buffer spaces over different time slots. Hence, we propose a simple heuristic algorithm, namely, the energy- and delay-aware network formation (EDA-NF) algorithm, to adapt the network formation based on the UAVs' energy consumption and buffer status.

The basic idea of the EDA-NF algorithm is to balance the energy consumption and queue size of different UAVs on the fly. Specifically, in each offloading sub-slot $t_{i,o}$, the UAV-$i$ also reports its current status to the BS, including its location ${\bm\ell}_i(t)$, current network formation $\Phi(t)$, energy consumption $\hat{e}_i(t)$, and the buffer {information $\lambda_i D_i(t)+\sum_{m\in\mathcal{M}_i} W_m(t)$, which} includes both the UAV-$i$'s buffer size $D_i(t)$ and all GUs' traffic demands $W_m(t)$ under the UAV-$i$'s coverage. When the BS collects all UAVs' status information, it will adapt the network formation $\Phi(t)$ to balance the UAVs' energy consumption and queue size. We assume that the UAVs' status information is of a small size and will not cause much overhead.

We first design a load balance coefficient $b_i(t)$ to characterize the UAV-$i$'s traffic conditions with respect to the overall network traffic, given as follows:
\begin{equation}\label{equ-balance}
b_i(t) = \frac{D_i(t)}{o_{i,0}(t)} -  \frac{1}{N-1} \sum_{j\neq i, j \in \mathcal{N} } \frac{D_j(t)}{o_{j,0}(t)},
\end{equation}
which depends on the UAVs' data size in buffers and the transmission capabilities via the U2B channels. The first term $\frac{D_i(t)}{o_{i,0}(t)}$ denotes the expected time delay when the UAV-$i$ forwards its data to the BS via the U2B channel. The second term in~\eqref{equ-balance} represents the average time delay of all other UAVs via their U2B channels. We can expect a large value $|b_i(t)|$ if the data traffic is unbalanced between the UAV-$i$ and the other UAVs, e.g., the UAV-$i$ has a relatively large data size in the buffer but with a poor U2B channel condition. In this case, instead of using the U2B channel, the UAV-$i$ needs to establish the U2U links with the nearby UAVs and offload a part of its data via the more preferable U2U channels. It is worth noting that the network formation may not need to be updated frequently in each time slot. In particular, the network formation will remain the same if all UAVs' load balance coefficients are relatively small.

Besides the load balance coefficient $b_i(t)$, we further define a cost function for each UAV-$i$ as follows:
\begin{align}\label{equ-cost}
c_i(t) = \hat{e}_i(t) + \lambda_i D_i(t) + \sum_{m=1}^M W_m(t),
\end{align}
which characterizes the UAV-$i$'s resource demands including its energy consumption $\hat{e}_i(t)$ and traffic demand. A larger cost value $c_i(t)$ implies that the UAV demands more energy and incurs excessive transmission delay due to unsatisfactory channel conditions for data offloading, while the coefficient $b_i(t)$ characterizes the direct U2B transmission capabilities. We first divide all UAVs into two groups according to the coefficient $b_i(t)$, denoted by the subsets $\mathcal{G}_1$ and $\mathcal{G}_2$, respectively. With a larger value $b_i(t)$, the UAV-$i$ has a relatively heavy workload and the unsatisfactory channel condition. In this case, we expect to establish the U2U link for the UAV-$i$ to offload its data to a neighboring UAV with a lighter workload. Hence, we can consider a threshold-based division between $\mathcal{G}_1$ and $\mathcal{G}_2$. When $b_i(t)$ is greater than a threshold $b_o$, the UAV-$i$ is aligned to the set $\mathcal{G}_1$ and otherwise to the set $\mathcal{G}_2$. Each UAV in the set $\mathcal{G}_1$ can establish U2U links with the other UAVs, while the UAVs in set $\mathcal{G}_2$ have direct U2B connections.

The idea of the EDA-NF algorithm is to establish the U2U connections between the UAVs in two subsets and thus allow the {heavy-loaded UAVs in $\mathcal{G}_1$ to offload data to the light-loaded UAVs} in $\mathcal{G}_2$. We can assign the U2U channel to the UAV-$i$ in the subset $\mathcal{G}_1$ with the highest cost value $c_i(t)$ and select the UAV-$j$ in the subset $\mathcal{G}_2$ with the smallest cost value $c_j(t)$ as the relay node of the UAV-$i$. We also need to ensure that the U2U channel from the UAV-$i$ to the UAV-$j$ should meet a minimum data rate requirement. The detailed procedures are listed in Algorithm~\ref{alg-eda-algorithm}.
{
Note that Algorithm~\ref{alg-eda-algorithm} relies on the network state information to estimate the values in~\eqref{equ-balance} and~\eqref{equ-cost}. The threshold value $b_o$ is also a critical design parameter, which can be estimated experimentally in the offline phase. The network information can be collected by the UAVs and forwarded to the BS along with their data transmissions. The UAVs can rely on a hand-shake protocols to establish the U2U links. The transmitting UAV can send the U2U request to the target UAV. If the target UAV is unwilling to serve as the relay node, the U2U link among them will not be feasible by sending back a negative response to the transmitting UAV.
}

\begin{algorithm}[t]
	\caption{Energy- and Delay-Aware Network Formation (EDA-NF) Algorithm}\label{alg-eda-algorithm}
	\begin{algorithmic}[1]
		\STATE Initialize network formation $\phi_{i,0} = 1$ for $i\in\mathcal{N}$
		\STATE Initialize $(c_i, b_i)$ for all UAVs
		\STATE Initialize $\mathcal{G}_1\leftarrow \emptyset$, $\mathcal{G}_2\leftarrow \mathcal{N}$
		\FOR {$i\in\{1,2,\ldots, N\}$}
		\STATE Update $b_i(t)$ and $c_i(t)$
		\STATE $\mathcal{G}_1\leftarrow\{i\}$ if $b_t > b_o$
		\STATE $\mathcal{G}_2\leftarrow\{i\}$ if $b_t \leq b_o$
        \ENDFOR
        \STATE Sort UAVs in $\mathcal{G}_1$ by the descent order of $c_i(t)$
        \STATE Sort UAVs in $\mathcal{G}_2$ by the ascent order of $c_i(t)$
        \FOR{any $i\in\mathcal{G}_1$ and $j\in\mathcal{G}_2$}
        \STATE Check the distance $d_{i,j}$ between UAV-$i$ and UAV-$j$
        \STATE If $d_{i,j}<d_k$ then $\phi_{i,j} \leftarrow 1$, $\phi_{i,0} \leftarrow 0$, $\phi_{j,0} \leftarrow 1$, and
        \STATE remove UAV-$i$ from $\mathcal{G}_1$ and remove UAV-$j$ from $\mathcal{G}_2$
        \ENDFOR
        \STATE Return the network formation strategy $\Phi(t)$
	\end{algorithmic}
\end{algorithm}



\subsection{Learning for Trajectory Optimization}

Given the network formation $\Phi(t)$, the second problem is to update the UAVs' trajectories {as follows}:
\begin{subequations}\label{prob-trajectory}
\begin{align}
\min_{\mathcal{L}_i} ~&~ \sum_{t=1}^T\left(\sum_{i=1}^N \Big( \hat{e}_i(t) + \lambda_i D_i(t) \Big )  + \sum_{m=1}^M W_m(t)\right)\label{obj-energy-new}\\
s.t. ~&~ \eqref{con-model} \text{ and } \eqref{con-uav-fly} - \eqref{con-uav-safe}.
\end{align}
\end{subequations}
The trajectory optimization in~\eqref{prob-trajectory} is high-dimensional and still complicated to solve directly due to the spatial-temporal interactions among different UAVs. In the sequel, we propose the model-free DRL method for problem~\eqref{prob-trajectory} by reformulating it into Markov decision process (MDP).

\subsubsection{MDP reformulation}
MDP can be simply characterized by a tuple $(\mathcal{S},\mathcal{A}, \mathcal{R})$, where $\mathcal{S}$ and $\mathcal{A}$ represent the state spaces and action spaces, respectively. The reward function $\mathcal{R}$ assigns each state-action pair $({\bf s}(t), {\bf a}(t))$ a quality evaluation. For a multi-UAV network, the state ${\bf s}(t)$ includes all UAVs' local states, i.e.,~${\bf s}(t) = ({\bf s}_1(t), {\bf s}_2(t), \ldots, {\bf s}_N(t))$. Each UAV's local state ${\bf s}_i(t)$ includes its location $\boldsymbol{\ell}_i(t)$, {network formation} $\{\phi_{i,j}(t)\}_{j\in\tilde{\mathcal{N}}}$, energy status $E_i(t)$, and buffer size $D_i(t)$. Similarly, we have ${\bf a}(t) = ({\bf a}_1(t), {\bf a}_2(t), \ldots, {\bf a}_N(t))$ as the joint actions for all UAVs. Each UAV's action ${\bf a}_i(t)$ includes the flying direction ${\bf d}_i(t)$ and speed $v_i(t)$ in each time step.

Each UAV-$i$ can obtain its reward $R_i({\bf s}(t), {\bf a}_i(t))$ when it takes an action ${\bf a}_i(t)$ on the state ${\bf s}(t)$ in $t$-th time slot. It is clear that the UAV-$i$'s reward also depends on the other UAVs' actions, denoted as ${\bf a}_{-i}(t)$. Specifically, the reward function $R_i({\bf s}(t), {\bf a}_i(t))$ can be characterized by three parts: the energy reward $R_{i,e}(t)$, the transmission reward $R_{i,d}(t)$, and the sensing reward $R_{i,c}(t)$. The energy reward is simply defined as the negative of the UAV-$i$'s energy consumption:
\[
R_{i,e}(t) = -\hat{e}_i(t),
\]
which urges the UAV-$i$ to reduce its energy consumption. To reduce the transmission delay, each UAV receives a reward if it forwards the data as much as possible. Hence, the transmission reward $R_{i,d}(t)$ is proportional to the size of successfully transmitted data to the BS or the next-hop UAVs, i.e.,
\[
R_{i,d}(t)  = o_{i,0}(t) + \sum_{j\in\mathcal{N}} o_{i,j}(t).
\]
The sensing reward is used to promote the UAVs to collect more data from the GUs. {Hence, we define the UAV-$i$'s sensing reward $R_{i,s}(t)$ as} the size of sensing data collected from the GUs under its coverage:
\begin{equation}\label{equ-sensing-data}
R_{i,s}(t) = \sum_{m\in\mathcal{M}_i} s_{i,m}(t).
\end{equation}
Besides the above reward terms, an additional penalty term $R_{i,p}(t)$ {is imposed} to ensure a minimum safe distance between the UAV-$i$ and other UAVs. We can simply assign {a larger} value to $R_{i,p}(t)$ if the constraint in~\eqref{con-uav-safe} does not hold, i.e.,
\[
R_{i,p}(t) = \mu \sum_{j\in\mathcal{N}} {\bf I}( ||\boldsymbol{\ell}_i(t) - \boldsymbol{\ell}_j(t) || < d_{\min} ),
\]
where ${\bf I}(\cdot)$ denotes an indicator function. To this point, we can use different combining weights $\gamma_i$ to define {the UAV-$i$'s overall reward function} in each time slot as follows:
\[
R_{i}({\bf s}(t),{\bf a}(t)) = \gamma_1 R_{i,e}(t) + \gamma_2 R_{i,d}(t) + \gamma_3 R_{i,s} (t)- R_{i,p}(t).
\]

\subsubsection{Multi-agent DRL} {The continuous control problem in~\eqref{prob-trajectory} can be flexibly handled by the actor-critic DRL framework using} two sets of deep neural networks (DNNs) to approximate the policy and value functions, respectively. Let ${\bm\theta}$ denote the DNN parameter of {the UAV's} policy function, namely the actor-network. Focusing on deterministic policy, the parameterized actor-network $\pi({\bf s}_t |{\bm\theta})$ will generate a deterministic action ${\bf a}_t = \pi({\bf s}_t |{\bm\theta})$ on each state ${\bf s}_t$ to maximize the value function defined as follows:
\begin{equation}\label{equ-value-func}
J({\bm \theta}) = \sum_{{\bf s} \in \mathcal{S}} d^{\pi}({\bf s}) Q^{\pi}({\bf s},{\bf a} ) \approx \mathbb{E}_{t\in \mathcal{B}}[ Q^{\pi}({\bf s}_t, \pi( {\bf s}_t | {\bm \theta}) ],
\end{equation}
where $d^{\pi}({\bf s})$ denotes the stationary state distribution following the deterministic policy ${\bf a}_t=\pi({\bf s}_t |{\bm\theta})$. Given a limited set of state transition samples, we can use the expectation over the sampling space $\mathcal{B}$ to approximate the value function $J({\bm \theta})$ {in~\eqref{equ-value-func}}. The Q-value $Q^{\pi}({\bf s}_t,{\bf a}_t)$ helps evaluate the quality of the policy $\pi({\bf s}_t |{\bm\theta})$, i.e.,~a large Q-value implies that the action ${\bf a}_t = \pi({\bf s}_t |{\bm\theta})$ can be more preferable when visiting the same state ${\bf s}_t$ in the future time steps. However, the true Q-value may not be available during online learning. We further {require} the critic-network with the DNN parameter ${\bm w}$ to approximate it, denoted as $Q^{\pi}({\bf s}_t,{\bf a}_t |{\bm w})$. As such, the value function in~\eqref{equ-value-func} depends on both {the DNN parameters ${\bm \theta}$ and ${\bm w}$ for the actor- and critic-networks, respectively}

{Taking the derivative of $J({\bm \theta})$} with respect to the policy parameter ${\bm \theta}$, we can update the actor-network by the gradient ascent direction to improve the value function $J({\bm \theta})$. By the deterministic policy gradient (DPG) theorem~\cite{maddpg17}, the policy gradient can be estimated as follows:
\begin{equation}\label{equ-pgradient}
\nabla J({\bm \theta}) = \mathbb{E}_{t\in \mathcal{B}}[ \nabla_{{\bf a}_t} Q^{\pi}({\bf s}_t, {\bf a}_t | {\bm w}) \nabla_{\bm \theta} \pi({\bf s}_t |{\bm\theta}) ],
\end{equation}
The policy gradient in~\eqref{equ-pgradient} relies on both the parameter gradients in the actor- and critic-networks, which can be easily evaluated by gradient {back-propagation method}. The critic-network can be updated by the temporal-difference (TD) error between the online critic-network $Q^{\pi}({\bf s}_t,{\bf a}_t |{\bm w})$ and its target $y_t = r_t + \gamma Q^{\pi}({\bf s}'_t,{\bf a}'_t |{\bm w}')$, where $r_t$ denotes the immediate reward and ${\bm w}'$ denotes {the DNN parameter} of the target critic-network. The training of the critic-network aims to minimize the TD error by the gradient descent direction:
\begin{equation}\label{equ-loss}
L({\bm w}) = \mathbb{E}_{t\in \mathcal{B}}[|y_t - Q^{\pi}({\bf s}_t, {\bf a}_t|{\bm w})|^2].
\end{equation}
To improve the stability in learning, both actor- and critic-networks have their target versions with the parameters ${\bm \theta}'$ and ${\bm w}'$, respectively, which can be updated from the online parameters $({\bm \theta}, {\bm w})$ smoothly~\cite{drl_survey}.

\begin{figure}[t]
	\centering
	\includegraphics[width=1\linewidth]{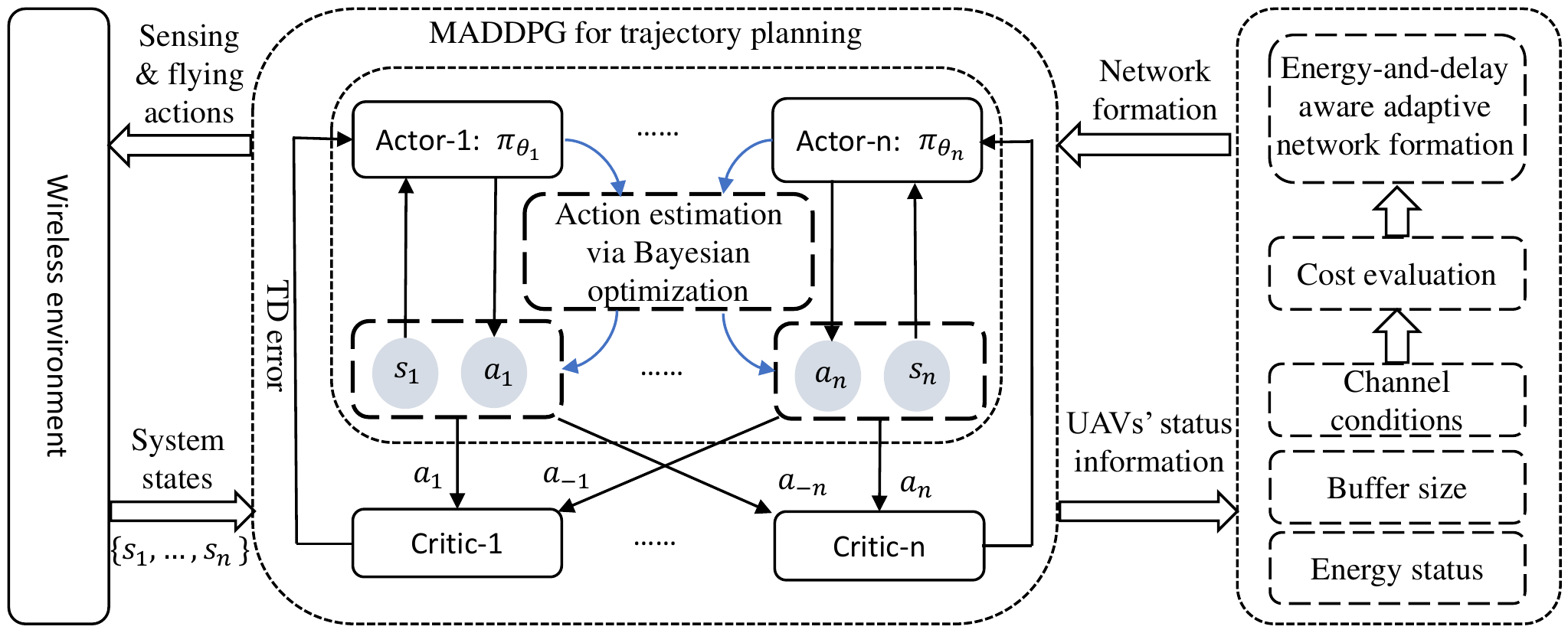}
	\caption{The learning framework for joint network formation and trajectory optimization.}
	\label{fig:Algorithm_Summary_simple}
\end{figure}

{In multi-UAV-assisted wireless networks}, each UAV's observation not only depends on its own action, but also relates to the other UAVs' actions. We can use the MADDPG algorithm to learn the UAVs' trajectories, by using the centralized training and decentralized execution scheme~\cite{maddpg17}. The above analysis in~\eqref{equ-value-func}-\eqref{equ-loss} needs to be revised slightly. Specifically, we assume that each UAV-$i$ is an independent DRL agent with the policy parameter ${\bm\theta}_i$, which outputs its own action ${\bf a}_i$ using the deterministic policy $\pi_i( {\bf o}_i | {\bm\theta}_i )$ based on its own observation ${\bf o}_i$ of the system. Note that the UAV-$i$'s observation ${\bf o}_i$ is not exactly the system state due to partial observability~\cite{ekram-madrl}. Each UAV-$i$ {also has its own Q-value estimation, which depends on} all UAVs' joint actions. Let ${\bf o}_{-i}$ and ${\bf a}_{-i}$ denote the observations and actions of the other UAVs, respectively. We can revise the UAV-$i$'s policy gradient in~\eqref{equ-pgradient} as follows:
\begin{equation}\label{equ-ma-pg}
\nabla_{{\bm\theta}_i} J({\bm \theta}_i) = \mathbb{E}_{\mathcal{B}}[ \nabla_{{\bf a}_i} Q_i^{\pi}( {\bf o}, {\bf a} | {\bm w}_i) \nabla_{{\bm \theta}_i} \pi_i({\bf o}_i |{\bm\theta}_i) ],
\end{equation}
where ${\bf o} \triangleq ({\bf o}_{i}, {\bf o}_{-i})$ and  ${\bf a} \triangleq ({\bf a}_{i}, {\bf a}_{-i})$ denote the joint observations and actions of all UAVs. The expectation is taken over all samples in the experience replay buffer $\mathcal{B}$.

The revision to individual UAV's policy gradient in~\eqref{equ-ma-pg} reveals that the actor-network $\pi_i({\bf o}_i |{\bm\theta}_i)$ can be localized while the critic-network $Q_i^{\pi}( {\bf o}, {\bf a} | {\bm w}_i)$ requires the global information $({\bf o}, {\bf a})$. This motivates the popular centralized training and decentralized execution scheme for multi-agent systems. The centralized training in offline phase requires the BS to collect all UAVs' status updates and train the critic- and actor-networks simultaneously. After offline training, the critic- and actor-networks can be announced to different UAVs and used to {guide their decision-making and action execution} in a decentralized manner. The algorithm framework of the joint network formation and trajectory optimization is shown in Fig~\ref{fig:Algorithm_Summary_simple}. Given the network formation $\Phi(t)$, each UAV-$i$ updates its trajectory $\mathcal{L}_i$ by searching for the optimal flying action in the next time step using the MADDPG algorithm. Then, each UAV follows its trajectory to collect the GUs' {sensing data} and forwards it to the BS or the next-hop UAV. Meanwhile, the UAV can report its {status update} to the BS. As such, the BS can examine the quality of {the current network formation} strategy $\Phi(t)$ by evaluating the UAVs' load balance coefficients and cost functions in~\eqref{equ-balance} and~\eqref{equ-cost}, respectively. If the current network formation exaggerates the network conditions, e.g.,~unbalanced energy consumption and buffer size, the adaptive network reformation will be required to {restore the} balance in resource consumption among UAVs.

\section{Action Estimation for MADDPG via Bayesian Optimization} \label{sec-bayesian}

The MADDPG algorithm provides a general solution framework for complex control problems in multi-agent systems. However, it is still challenging to apply {it directly to} the multi-UAV trajectory optimization problem. Firstly, it requires collecting all UAVs' observations and then adapting their flying actions jointly. Each UAV needs to report its local observation to the BS, including the channel information, energy status, and the queue size in its buffer. This can be problematic in the fast-changing UAV network due to the transmission {delay in} multi-hop relay communications. The global information at the BS for centralized training can become obsolete due to the UAVs' mobility. Besides, the state and action spaces increase rapidly as the number of UAVs increases. The communication overhead for exchanging the UAVs' local observations and actions also becomes significant. This incurs excessive training overhead and leads to {instability in convergence}.

\begin{figure}[t]
	\centering
	\includegraphics[width=0.7\linewidth]{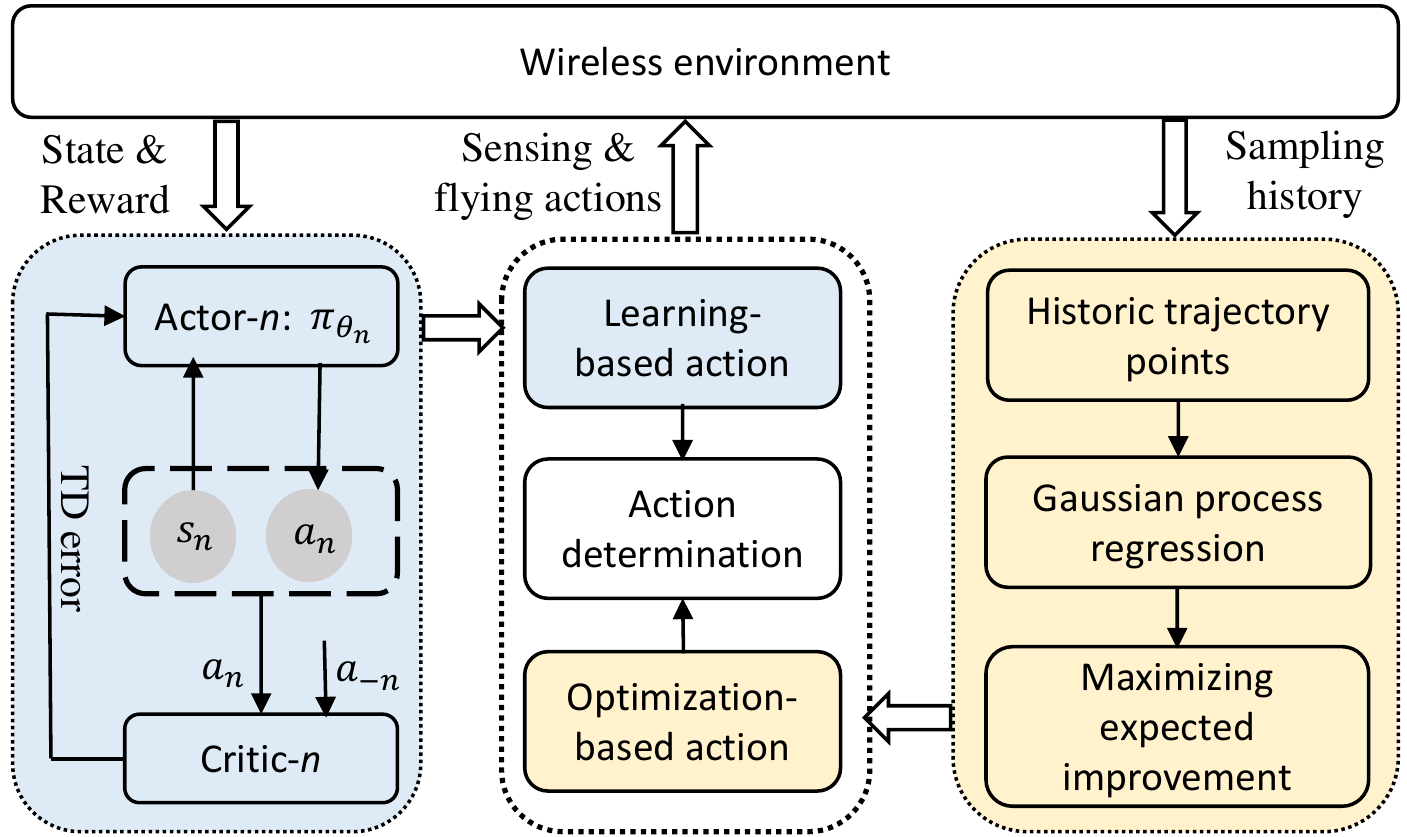}
	\caption{Bayesian optimization enhanced MADDPG (BO-MADDPG).}
	\label{fig-bayesian}
\end{figure}

In this part, we aim to improve the learning efficiency of MADDPG by using Bayesian optimization to estimate the UAVs' optimal flying actions in the next time step~\cite{tut-bo-2020}. Given the UAVs' observations along the past trajectories, Bayesian optimization provides a general mathematical framework for model-free prediction based on existing sampling data. The basic idea is that the action estimation via Bayesian optimization can provide {a better-informed direction of action exploration}, compared to the random action exploration. Therefore, it can guide the UAVs' trajectory learning towards a more rewarding policy. Such a guided learning can be viewed as semi-supervised learning by providing the action estimation to the DRL agent. This can be very useful in the early stage of multi-agent learning, where the random action exploration may need a {large number of learning episodes to warm up}.

The Bayesian optimization enhanced MADDPG framework (denoted as BO-MADDPG) can be divided into three parts as shown in Fig.~\ref{fig-bayesian}. The action estimation module takes the UAVs' most recent trajectory points as input and estimates the optimal flying locations in the next time step by using the Bayesian optimization method. The action estimation will be further input to the critic-network together with the action learned by the actor-network. Then the critic-network evaluates the qualities of two actions based on the current observations. The critic's quality evaluation helps decide which action will be executed in the network environment. Typically, we can choose a more rewarding action with a higher probability and avoid fruitless action exploration. This can potentially reduce the action space and improve the multi-agent learning efficiency.



\subsection{Action Estimation via Bayesian Optimization}
The MADDPG algorithm adapts the UAVs' trajectory points by trial-and-error exploration, which can be very inefficient especially in the early stage of learning. Bayesian optimization can be used to estimate the UAVs' trajectory points in a simpler and more efficient way. Specifically, we first decompose the multi-UAV trajectory planning problem into $N$ single-UAV trajectory planning problems based on local observations. Then, we can perform Bayesian optimization for each UAV to estimate its flying action in the next time step. 

\subsubsection{Data-driven probabilistic model}

For each UAV-$i$, let ${\bm\ell}_i(t)$ denote its location and $s_{i}(t) = \sum_{m\in\mathcal{M}_i} s_{i,m}(t)$ denote the sensing data collected by the UAV-$i$ in the $t$-th time slot, where $s_{i,m}(t)$ is given by~\eqref{equ-g2u}. Define a function as follows:
\[
f_i: {\bm\ell}_i(t)\rightarrow s_{i}(t) + \epsilon_i(t)
\]
to map each location point ${\bm\ell}_i(t)$ to the data size $s_{i}(t)$ collected from the GUs. Note that the function value $f_i({\bm\ell}_i(t))$ only provides an approximation to the sampling data $s_{i}(t)$. The error term $\epsilon_i(t)$ between $s_{i}(t)$ and $f_i({\bm\ell}_i(t))$ can be regarded as independent, identically distributed Gaussian noise with zero mean. By using Bayesian optimization, we aim to build a probabilistic model $\mathcal{P}({ f}_i(\mathcal{H}_t) |\mathcal{D}_i(t))$ based on a set of historical sampling points, denoted as $\mathcal{D}_i(t)\triangleq ({\bm\ell}_i(\tau), s_{i}(\tau))_{\tau\in\mathcal{H}_t}$, where $\mathcal{H}_t=\{ t-t_o,\ldots, t-1, t\}$ represents a set of time slots in the past. {The model} $\mathcal{P}({ f}_i(\mathcal{H}_t) |\mathcal{D}_i(t))$ represents the posterior probability distribution of the function values ${ f}_i(\mathcal{H}_t) \triangleq \{f_i(\tau)\}_{\tau\in\mathcal{H}_t}$ at different trajectory points $\{{\bm\ell}_i(\tau)\}_{\tau\in\mathcal{H}_t}$. It is clear that a larger size $|\mathcal{D}_i(t)|$ of historical samples can provide more information for accurate estimation {of the posterior distribution $\mathcal{P}({ f}_i(\mathcal{H}_t) |\mathcal{D}_i(t))$}. By Bayes' theorem, the posterior distribution given $\mathcal{D}_i(t)$ relates to the prior distribution of ${ f}_i(\mathcal{H}_t)$ and the likelihood {function $\mathcal{P}(\mathcal{D}_i(t)| { f}_i(\mathcal{H}_t) )$}:
\begin{equation}\label{equ-poster}
{ f}_i(\mathcal{H}_t) |\mathcal{D}_i(t)  \sim  \mathcal{P}(\mathcal{D}_i(t)| { f}_i(\mathcal{H}_t) ) \mathcal{P}({ f}_i(\mathcal{H}_t)).
\end{equation}
Since the UAVs are unaware of the GUs' spatial distribution and their traffic demands, we can use multi-variant Gaussian distribution $\mathcal{G}$ to model the prior distribution $\mathcal{P}({ f}_i(\mathcal{H}_t))$~\cite{tut-bo-2020}:
\begin{equation}\label{equ-gp}
{ f}_i(\mathcal{H}_t) \sim \mathcal{G}({\bm \mu}_i({\mathcal{H}_t}),{\bm V}_i(\mathcal{H}_t)),
\end{equation}
where ${\bm \mu}_i({\mathcal{H}_t})$ is the mean vector and ${\bm V}_i(\mathcal{H}_t)\triangleq\{v_{\tau,\tau'}\}_{\tau,\tau'\in\mathcal{H}_t}$ is the covariance matrix or the kernel function for each sampling value $s_{i}(\tau)$ on the trajectory point ${\bm\ell}_i(\tau)$ for $\tau\in\mathcal{H}_t$~\cite{gp-ml-book}. Initially, without any prior information we can assume zero mean ${\bm\mu}_i = {\bf 0}$. Besides, each element $v_{\tau,\tau'}$ of the covariance matrix can be defined as follows:
\begin{equation}\label{kernal_function}
v_{\tau,\tau'}(\mathcal{H}_t) = \exp\left(-\frac{1}{2} ||{\bm\ell}_i(\tau) - {\bm\ell}_i(\tau')||^2\right),
\end{equation}
which implies an intuitively larger correlation value $v_{\tau,\tau'}$ when two trajectory points ${\bm\ell}_i(\tau)$ and ${\bm\ell}_i(\tau')$ are closer to each other. Given the prior distribution in~\eqref{equ-gp} and the Gaussian likelihood $\mathcal{P}(\mathcal{D}_i(t)| { f}_i(\mathcal{H}_t) )$, we can easily obtain the posterior distribution $\mathcal{P}({ f}_i(\mathcal{H}_t) |\mathcal{D}_i(t))$ in~\eqref{equ-poster}, which can be recognized as a Gaussian distribution with the known mean and variance~\cite{gp-ml-book}.

\subsubsection{Predict the optimal trajectory point}
Now, we aim to predict the function value at a new trajectory point ${\bm\ell}_i(t+1)$ for the UAV-$i$, denoted as $f_i(t+1)$. Given the sampling history $\mathcal{D}_{i}(t)$, we can update the posterior distribution as follows:
\begin{equation}\label{equ-poster-new}
{f}_i(t+1)| \mathcal{D}_{i}(t) \sim \mathcal{G}({\mu}_i(t+1),\sigma^2_i(t+1)),
\end{equation}
where the mean and variance $({\mu}_i(t+1),\sigma^2_i(t+1))$ can be updated from~$({\bm \mu}_i({\mathcal{H}_t}),{\bm V}_i(\mathcal{H}_t))$. More detailed derivations can be referred to~\cite{tut-bo-2020} and Chapter 2.1 in~\cite{gp-ml-book}. With the increase in the size $|\mathcal{D}_{i}(t)|$, the posterior distribution of ${f}_i(t+1)$ will approach the true distribution. This allows us to evaluate the function values at different trajectory points.

{Aiming to collect more data from the GUs, each UAV-$i$ can} choose the next flying trajectory point ${\bm\ell}_i(t+1)$ to maximize the expected function value ${f}_i(t+1)$. Specifically, we define function $z_{i,t}({\bm\ell}_i)$ to characterize the expected improvement of the function value ${f}_i({\bm\ell}_i)$ as the UAV-$i$ moves to the trajectory point ${\bm\ell}_i$ at the $t$-th time slot:
\begin{equation}\label{equ-ei}
z_{i,t}({\bm\ell}_i) = \mathbb{E}[ \max\{0, {f}_i({\bm\ell}_i) - {f}_i^*(\mathcal{D}_i(t)) \}],
\end{equation}
where ${f}_i^*(\mathcal{D}_i(t))$ denotes the maximum function value in the past sampling points, i.e.,~${f}_i^* = \max_{{\bm\ell}_i \in\mathcal{D}_i(t)}f_i({\bm\ell}_i)$. Hence, the UAV-$i$'s optimal trajectory point in the next time slot will be found by maximizing the expected improvement, i.e.,~${\bm\ell}_i({t+1}) = \arg\max z_{i,t}({\bm\ell}_i)$. Practically, the search for the optimal trajectory point is also confined by the UAVs' speed and {range limits}. Compared to the random action exploration in MADDPG, the Bayesian estimation for the new trajectory point ${\bm\ell}_i(t+1)$ can be more informative as it motivates the UAV to collect more data. After moving to the new location, the UAV will collect the GUs' sensing data again and evaluate the quality of the network formation.

\subsection{Action Estimation for MADDPG}
The {Bayesian optimization enhanced MADDPG (BO-MADDPG)} framework is shown in Algorithm~\ref{alg-total-algorithm}. Firstly, the actor networks of different UAVs will generate their flying actions based on individuals' local observations. Meanwhile, the Bayesian optimization module of each UAV estimates the posterior distribution $\mathcal{P}({f}_i(t+1)| \mathcal{D}_{i}(t))$ based on historical sampling points $\mathcal{D}_i(t)$. By such a probabilistic model, the next flying trajectory point ${\bm\ell}_i(t+1)$ will be estimated by maximizing the expected improvement of the function value ${f}_i(t+1)$. {This encourages} each UAV to collect the sensing data as much as possible. Then, each UAV can fly from the current location ${\bm\ell}_i(t)$ to the next location ${\bm\ell}_i(t+1)$. For notational convenience, we denote the {actor-networks'} action predictions as $\{{\bf a}_1,{\bf a}_2,\ldots, {\bf a}_N\}$, and denote the location estimations by using Bayesian optimization as $\hat{\bf a}_i(t)$. All UAVs' action vector can be expressed as $\{\hat{\bf a}_1(t),\hat{\bf a}_2(t),\dots,\hat{\bf a}_N(t)\}$. For each UAV-$i$, {the critic-network will evaluate the qualities of both actions} ${\bf a}_1$ and $\hat{\bf a}_1$, denoted as ${q}_i$ and $\hat{q}_i$, respectively. The reference value $\hat{q}_i$ can be easily obtained based on the solution to the maximization problem in~\eqref{equ-ei}. A comparison between ${q}_i$ and $\hat{q}_i$ will decide the preference of the final trajectory point $\{{\bf a}^{*}_1,{\bf a}^{*}_2,\ldots , {\bf a}^{*}_N\}$ for all UAVs. A simple implementation is to take the action with a higher action-value, i.e.,~the UAV-$i$ will fly to the trajectory point ${\bm\ell}_i(t+1)$ if $\hat{q}_i$ is greater than ${q}_i$.

The detailed procedures are listed in Algorithm~\ref{alg-total-algorithm}. The two-step framework includes the adaptive network formation based on each UAV's cost parameters $(b_i(t), c_i(t))$ and the {multi-agent} trajectory planning module enhanced by Bayesian optimization algorithm. In lines $6-10$ of Algorithm~\ref{alg-total-algorithm}, with fixed network formation each UAV-$i$ decides the next trajectory point by using the actor-network and Bayesian optimization. Given two actions $({\bf a}_i, \hat{\bf a}_i)$ and the action-value estimations $(q_i, \hat q_i)$, the UAV-$i$ can simply take the greedy policy by always taking a higher action-value. Then, each UAV-$i$ executes the greedy action ${\bf a}^{*}_i$ and records the transition to the next state, as shown in lines $11-15$ of Algorithm~\ref{alg-total-algorithm}. After this, each UAV-$i$ evaluates its traffic conditions and cost values, and {then updates} the network formation by using Algorithm~\ref{alg-eda-algorithm}. {For practical implementation, we can rely on both the computation resources at the BS and individual UAVs. The BS can initialize the system and pre-train the UAVs' actor- and critic-networks. After that, each UAV can generate its local action based on local observation. Besides, each UAV can run the Bayesian optimization module to help estimate a preferable trajectory point. Note that the Bayesian optimization is not necessarily executed in each time step to minimize the UAVs' computational demands and foster practical implementation.}

\begin{algorithm}[t]
	\caption{BO-MADDPG Algorithm for multi-UAV Trajectory Planning}\label{alg-total-algorithm}
	\begin{algorithmic}[1]
        \STATE Initialize network formation $\Phi(t)$ and trajectories $\mathcal{L}_i(t)$
        \STATE Initialize observations $\mathcal{D}_i$ for all UAVs
        \FOR {Episode = $\{1,2,\ldots, \text{MAX}= 200K \}$}
        \STATE Each UAV constructs its state ${\bf s}_i(t)$.
        \FOR {each UAV agent $i\in\{1,2,\ldots, N\}$}
        \STATE Estimate posterior ${f}_i(t+1)| \mathcal{D}_{i}(t)$ based on historical sampling points $\mathcal{D}_i(t)$ using Bayesian optimization
        \STATE Maximize the expected improvement in~\eqref{equ-ei}
        \STATE Obtain the next trajectory point ${\bm\ell}_i(t+1)$
        \STATE Convert ${\bm\ell}_i(t+1)$ into the UAV's action $\hat{\bf a}_i(t)$
        \STATE Actor network updates action ${\bf a}_i(t) = \pi_i( {\bf o}_i | {\bm\theta}_i )$
        \STATE Evaluate the action-value $(q_i, \hat q_i)$ for $({\bf a}_i, \hat{\bf a}_i)$
        \STATE Take the action ${\bf a}^{*}_i$ with a higher action-value
        \STATE Record state transition $s_i(t+1)$ and the reward $R_{i}(t)$
        \STATE Store the transition and update the history $\mathcal{D}_i(t)$
        \STATE Update the UAV's actor- and critic-network
        \ENDFOR
        \STATE $t \leftarrow t+1 $
        \STATE Update $b_i(t)$ and $c_i(t)$ for all UAVs
        \STATE Update network formation $\Phi(t)$ by Algorithm \ref{alg-eda-algorithm}
        \ENDFOR
	\end{algorithmic}
\end{algorithm}

\section{Simulation Results}\label{sec-simulation}
In this part, we present simulation results to verify the performance gain by the joint network formation and trajectory optimization for multiple UAVs. A few GUs are distributed in the 2$\times$2 km$^2$ area as shown in Fig.~\ref{fig-model}. We scale the x-y coordinates to the range of $[-1, 1]$. The BS is far away from the service area and located in the upper right corner of {the service area}. Both the UAVs and the GUs have a fixed {transmit power} at 23 dBm. The UAVs' energy consumption model refers to that in~\cite{zeng2017energy}. More detailed parameters are listed in Table~\ref{tab-para}, similar to the parameter settings in~\cite{zhang2019cellular}.

\begin{table}[t]
    \centering
    \caption{Parameter settings in the numerical simulations.}
    \begin{tabular}{|l|l|}
    \hline
    Parameters & Settings \\ \hline
    Number of channels $K$ & 3 \\
    Path-loss coefficient $\alpha$  & 2 \\
    Altitude of UAVs $H$ & 100 m \\
    Maximum UAV speed $v_{\max}$  & 20 m/s\\
    Channel bandwidth & 1MHz \\
    GUs' data size $D_m$ & 10 M bits \\
    Noise power $\delta^2_k$ & $-90$ dBm \\
    Carrier frequency $f_c$  & 2 GHz\\
    Training cycles per episode  & 200 \\
    Sampling batch size & 256\\
    Actor's learning rate & $10^{-3}$ \\
    Critic's learning rate & $10^{-4}$\\
    $\epsilon$-greedy parameter & $0.1$ \\
    Noise rate for action exploration & 0.1\\\hline
    \end{tabular}
    \label{tab-para}
\end{table}

\subsection{Bayesian Optimization Improves Trajectory Planning}

\begin{figure}[t]
\centering
\subfloat[Bayesian optimization]{\includegraphics[width=0.45\linewidth]{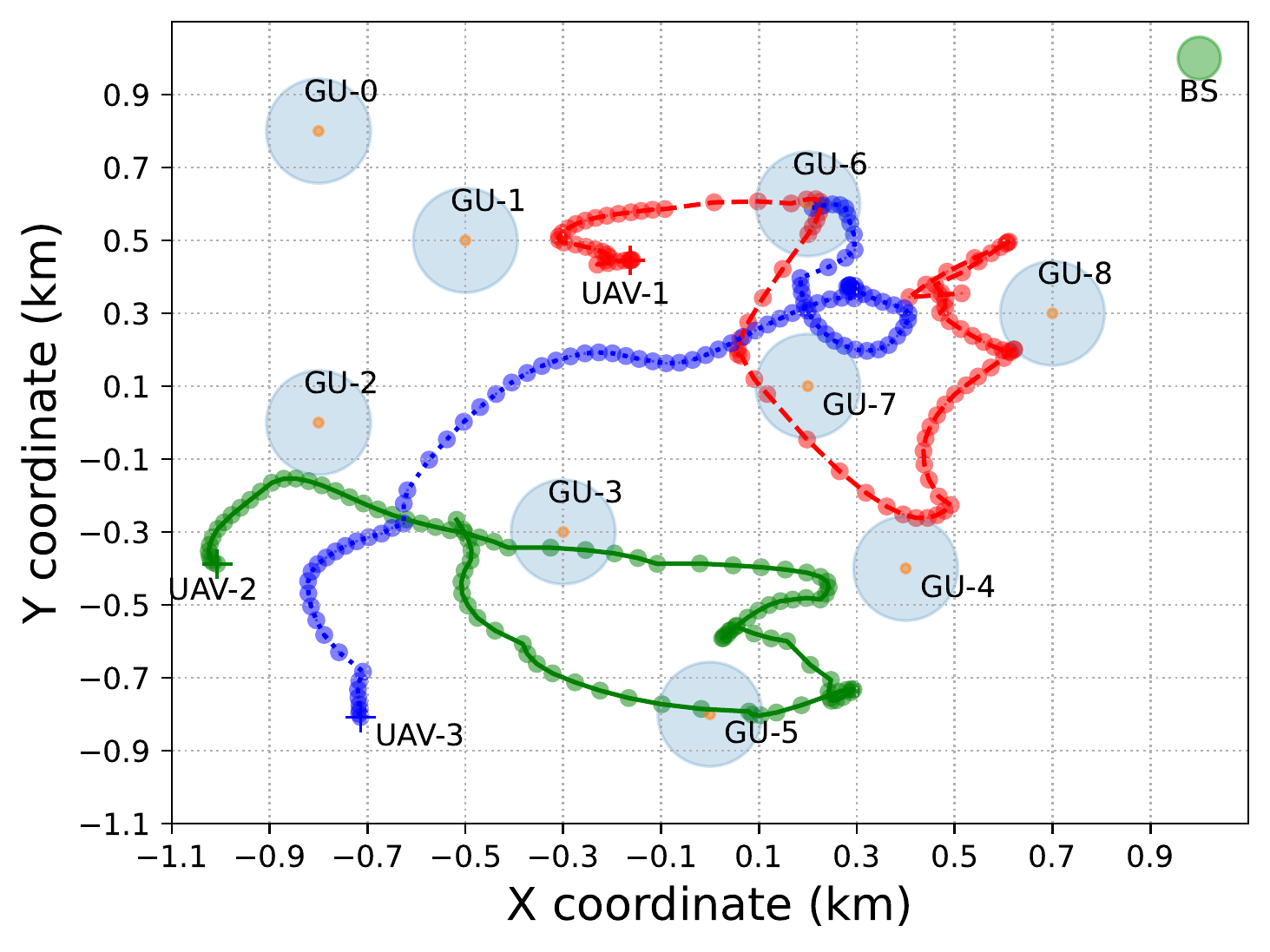}}
\subfloat[Layered-MADDPG]{\includegraphics[width=0.45\linewidth]{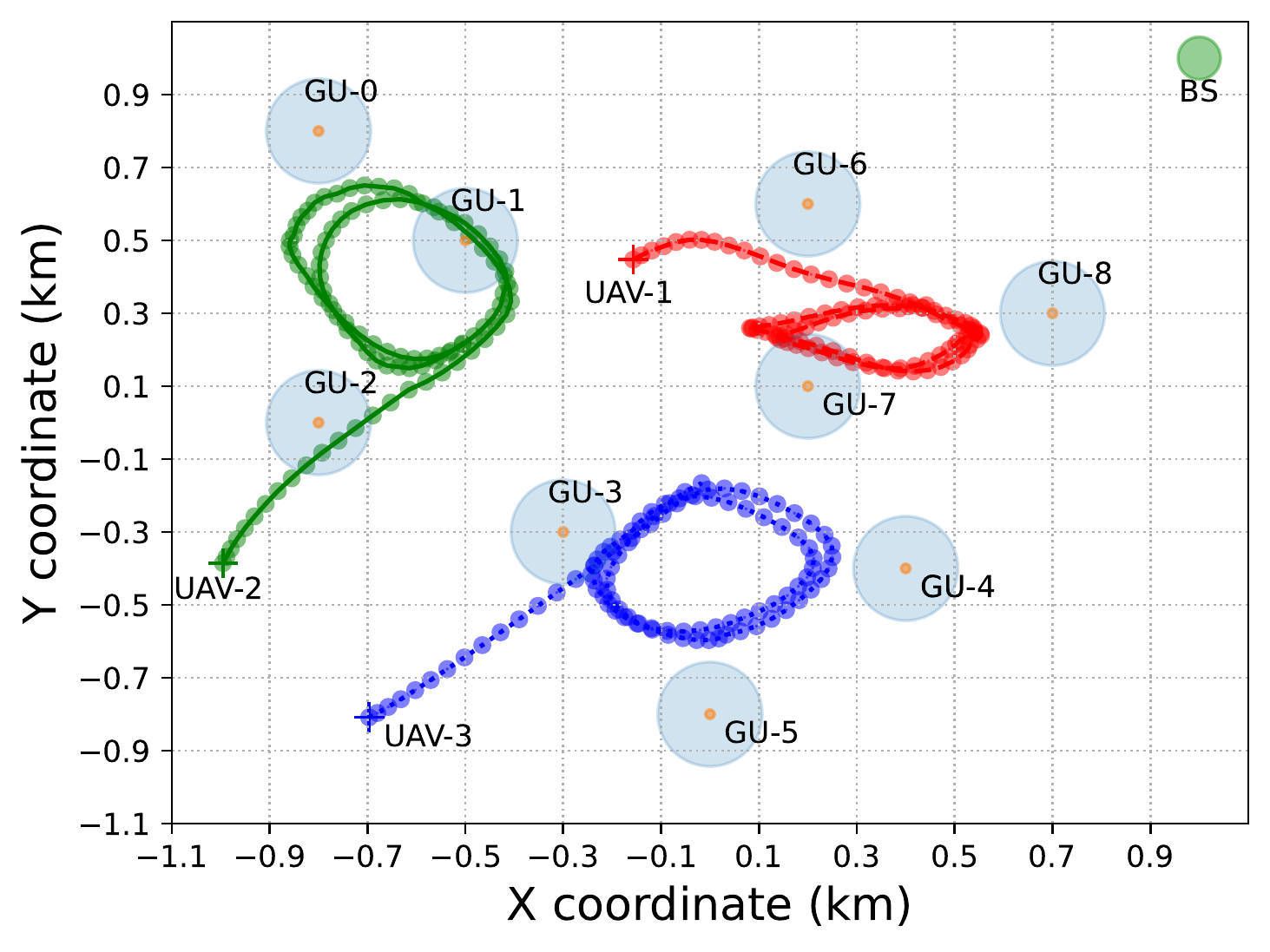}}\\
\subfloat[BO-MADDPG (Case I)]{\includegraphics[width=0.45\linewidth]{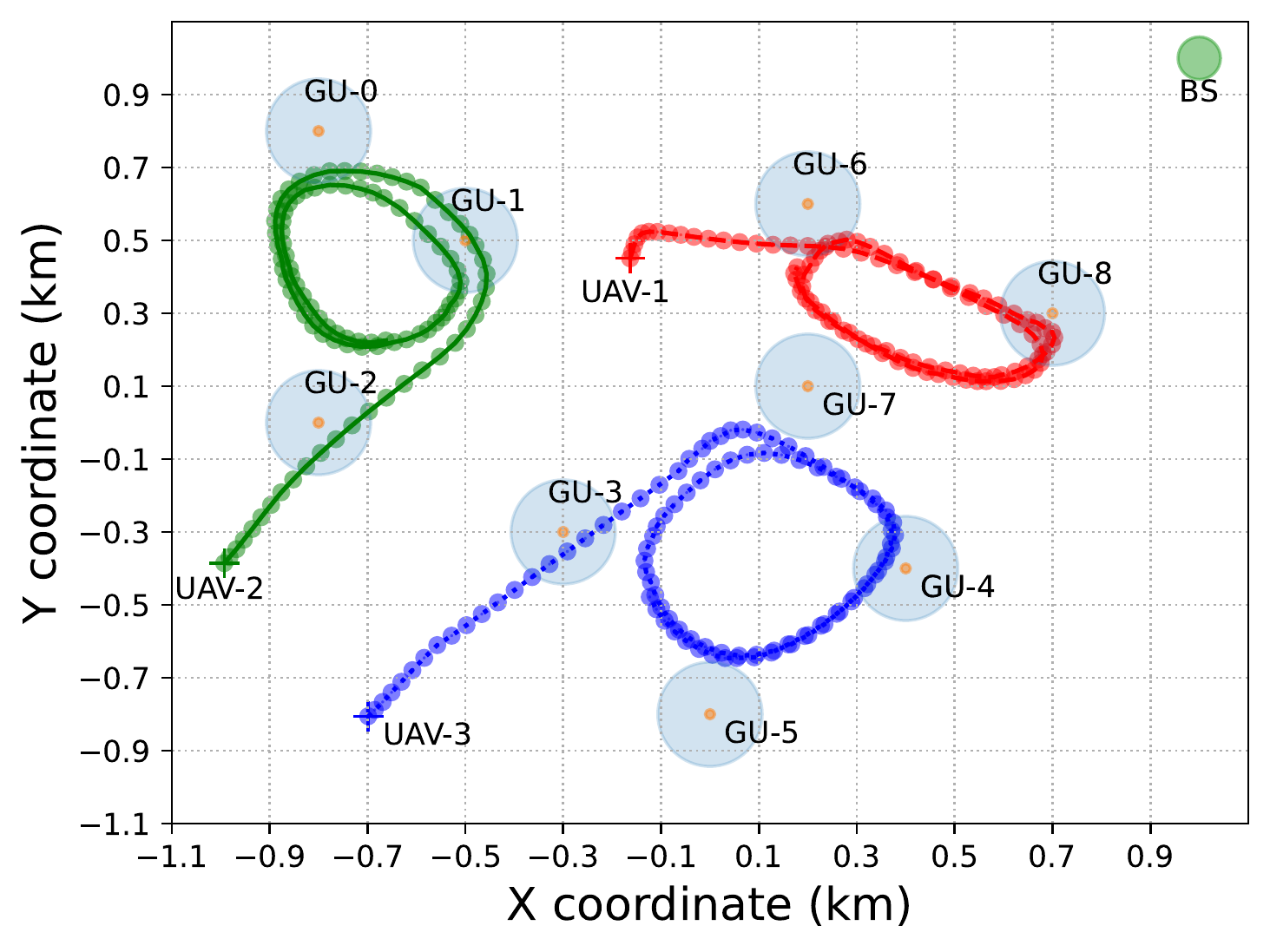}}
\subfloat[BO-MADDPG (Case II)]{\includegraphics[width=0.45\linewidth]{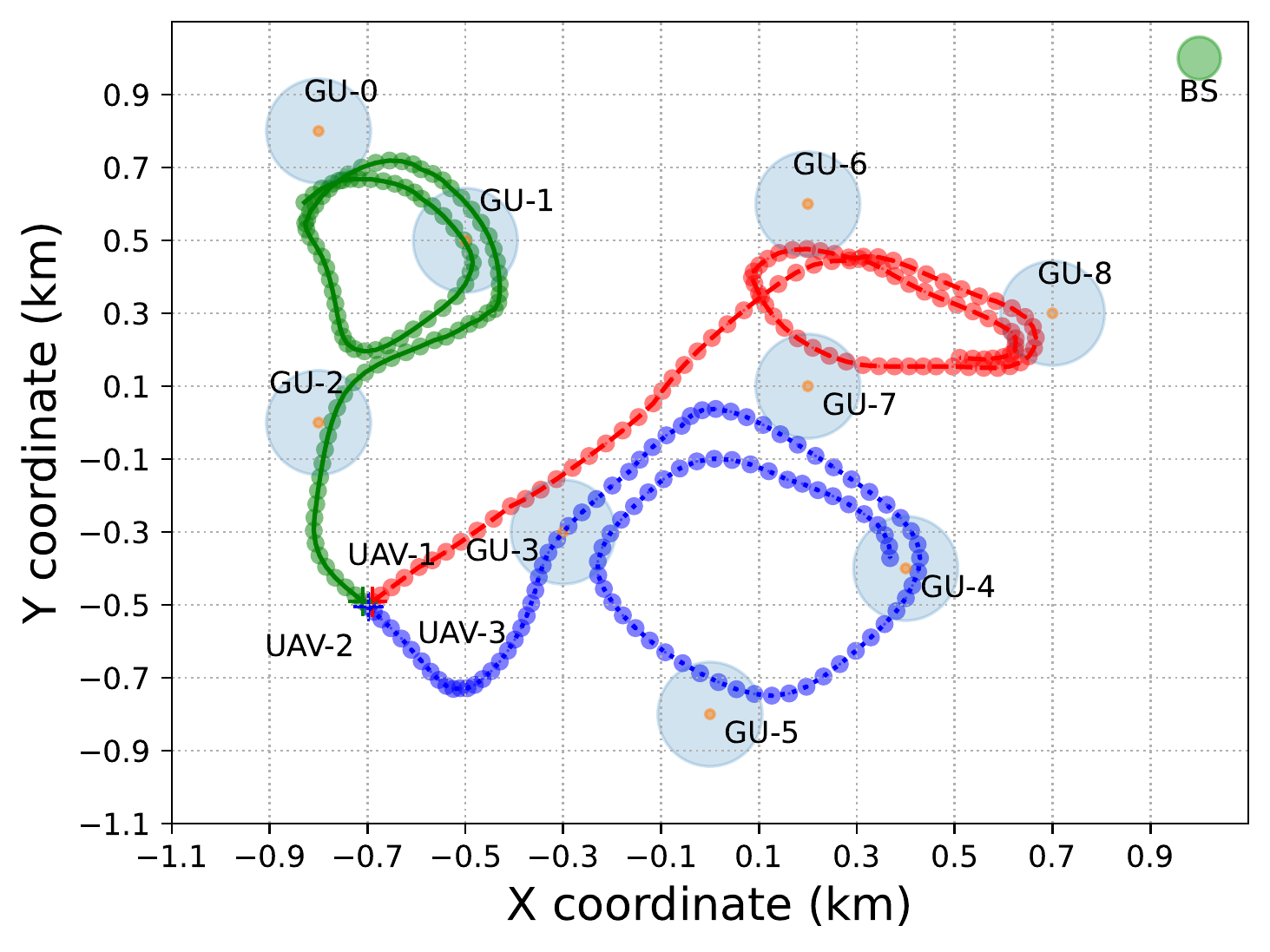}}
\caption{The evolution of UAVs' trajectories in different algorithms.}
\label{fig:Compare}
\end{figure}

{In our algorithm design, we first propose the Layered-MADDPG framework that relies on the EDA-NF algorithm to adapt the network formation while using the conventional MADDPG algorithm to learn the UAVs' trajectories, thereby reducing the agent's exploration action space. Besides, we further improve the MADDPG algorithm by using the Bayesian optimization module estimate the network environment and thus improve the trajectory learning efficiency, which is denoted as the BO-MADDPG method.} Fig.~\ref{fig:Compare} visualizes the UAVs' trajectories in different algorithms, including the Bayesian optimization algorithm, {the Layered-MADDPG, and the BO-MADDPG, which is built on the Lyaered-MADDPG and the Bayesian optimization methods}. Different colors in Fig.~\ref{fig:Compare} indicate the flying locations of different UAVs. Each UAV takes off from a random starting point and collects the sensing data from the GUs along its trajectory. Fig.~\ref{fig:Compare}(a) shows the trajectory planning result guided by Bayesian optimization method. In this case, each UAV estimates its best flying location in the next time slot based on the historical trajectory points of its own. It does not take into account the task cooperation between different UAVs. As a result, the performance of Bayesian optimization method is limited in a decentralized multi-UAV network. The lack of coordination among different UAVs may lead to service confliction or resource wastage. For example, different UAVs may have overlapped service range.

Fig.~\ref{fig:Compare}(b) shows the trajectory planning by using the {Layered-MADDPG algorithm}, which makes a trajectory policy for each UAV based on the centralized training and decentralized execution scheme. The centralized training requires global information of all UAVs. After learning, multiple UAVs can cooperatively divide the entire area into different parts and each part of the GUs can be served by a different UAV. We can observe that the whole sensing task is finally divided into three separated task groups. Each task group contains a subset {of nearby GUs}. Then, each UAV will focus on the data collection of one task group by flying around these GUs. Such a spatial division of the sensing task can better exploit the UAVs' cooperation and thus potentially reduce the overall energy consumption and transmission delay. Similar to the {Layered-MADDPG} in Fig.~\ref{fig:Compare}(b), the BO-MADDPG also divides the GUs into different groups as shown in Fig.~\ref{fig:Compare}(c). However, the group-edge GUs (e.g., the GU-7) are given {higher priorities and served with higher transmission rates}. Besides, the trajectory planning in BO-MADDPG can be more stable comparing with the {Layered-MADDPG} algorithm. Different from Fig.~\ref{fig:Compare}(c), we further assume that all UAVs start from the same initial location as shown in Fig.~\ref{fig:Compare}(d). After learning, the UAVs can spread out to serve different areas. This shows the efficacy of the BO-MADDPG algorithm to exploit the multi-UAV's cooperation. 

\begin{figure}[t]
\centering
\subfloat[BO-MADDPG (Case I) ]{\includegraphics[width=0.45\linewidth]{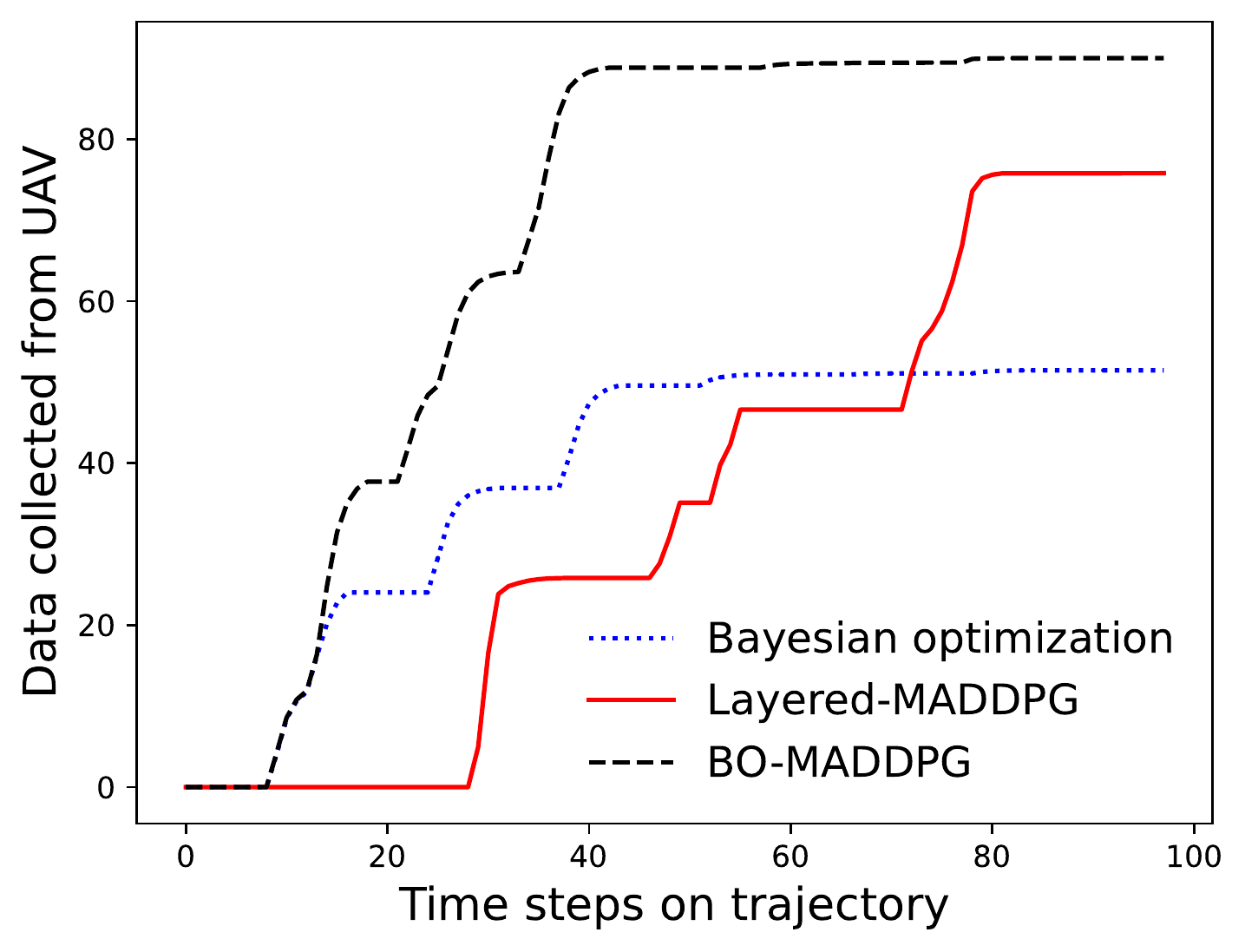}}
\subfloat[BO-MADDPG (Case II)]{\includegraphics[width=0.45\linewidth]{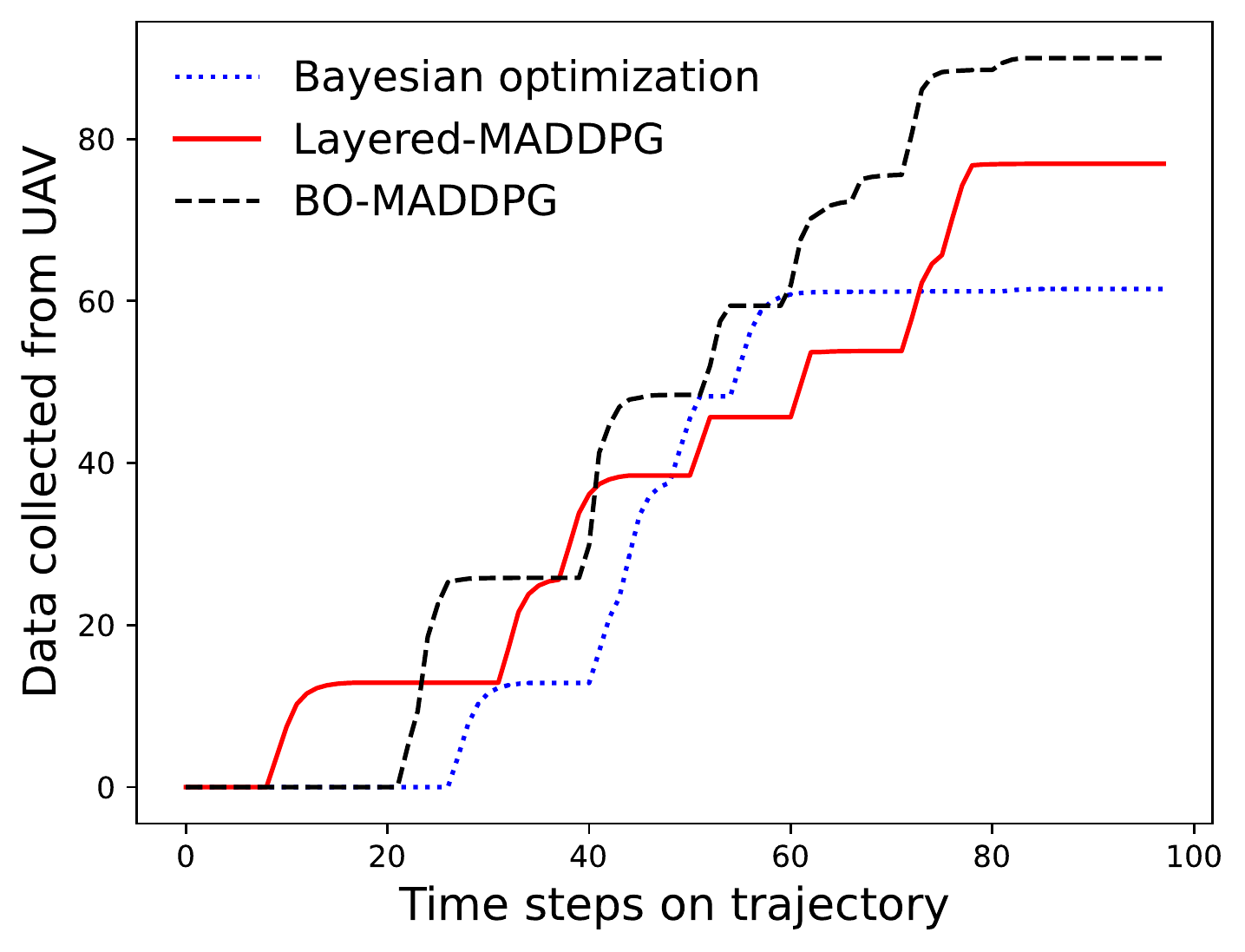}}
\caption{The size of data collected from UAVs in different time steps.}
\label{fig:rew_compare}
\end{figure}

Fig.~\ref{fig:rew_compare} shows the size of sensing data collected by different UAVs along their trajectories. In each time slot $t$, the sensing data received by the UAV-$i$ is given by $R_{i,s}(t)$, as defined in~\eqref{equ-sensing-data}. The comparison in Fig.~\ref{fig:rew_compare}(a) shows that the BO-MADDPG algorithm achieves a larger quantity of sensing data comparing to the other two cases. This verifies the advantages of our design concept by using partial information to guide the model-free learning approach towards a much better reward. Fig.~\ref{fig:rew_compare}(b) shows a similar result when the UAVs are given the same initial location corresponding to the trajectories in Fig.~\ref{fig:Compare}(d). For either case, the BO-MADDPG algorithm can collect the most sensing data from the GUs.

\begin{figure}[t]
\centering
\subfloat[Better reward performance]{\includegraphics[width=0.45\linewidth]{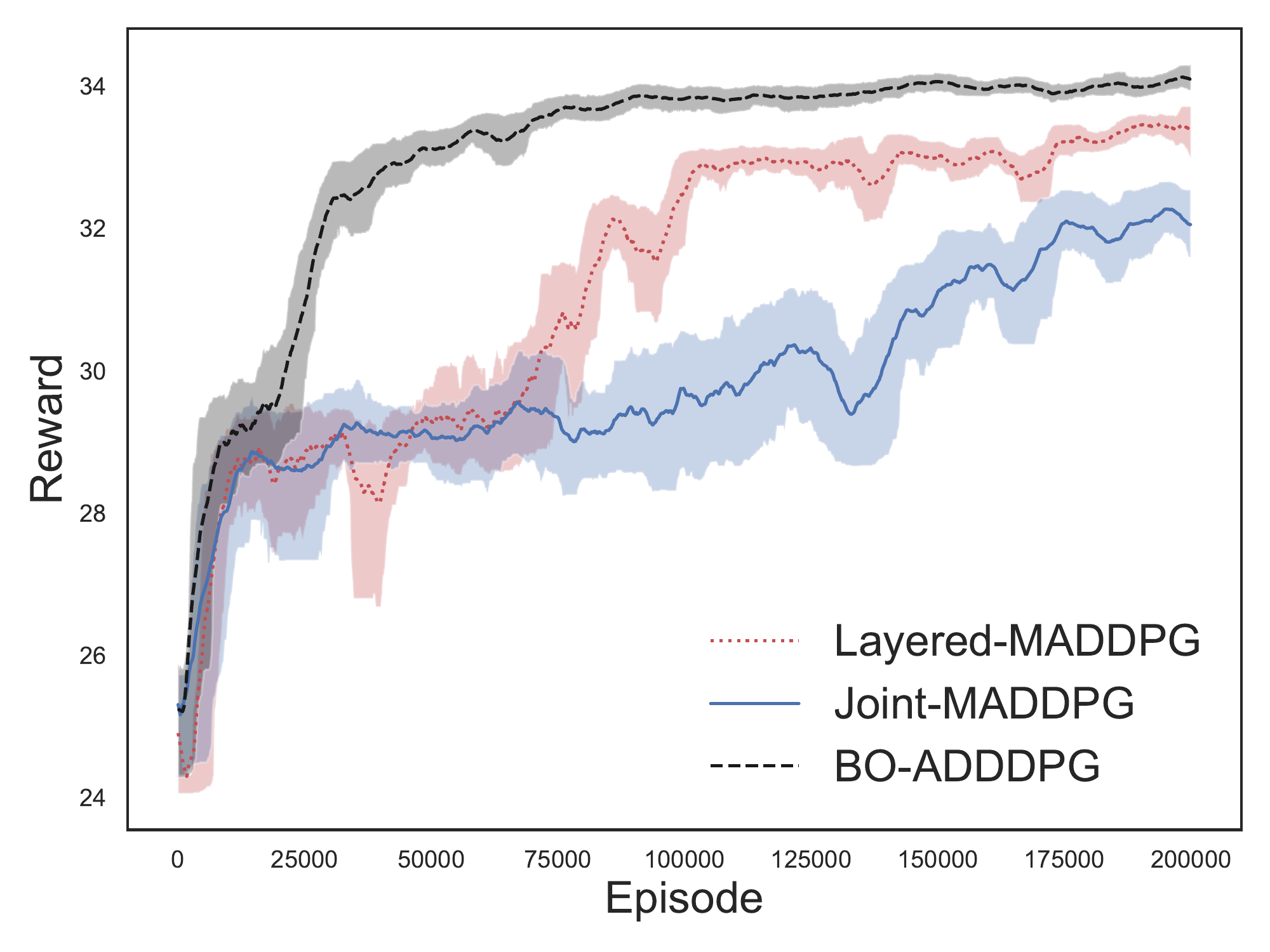}}
\subfloat[More stable learning]{\includegraphics[width=0.45\linewidth]{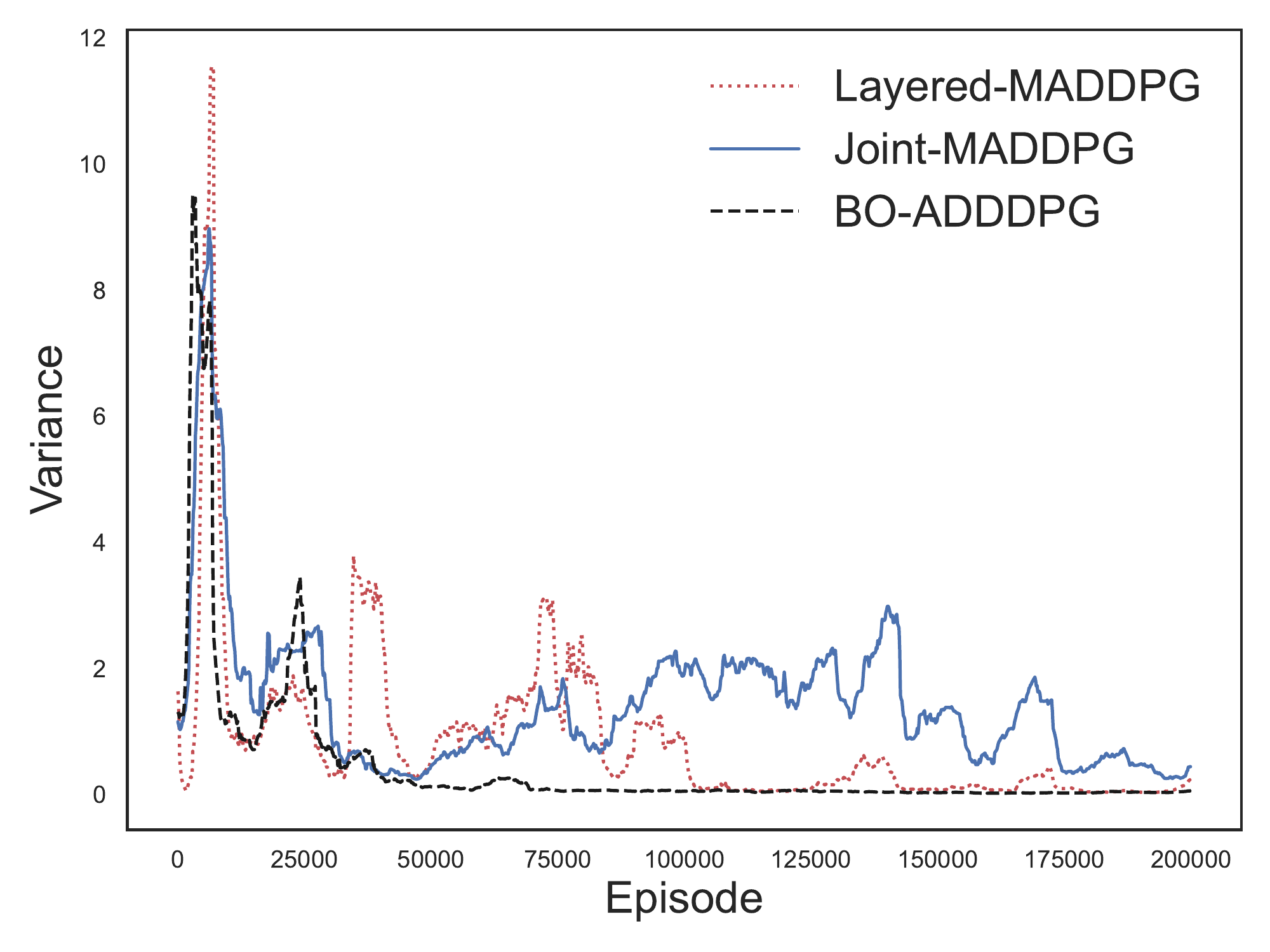}}
\caption{Convergence performance of different MADDPG algorithms.}
\label{fig:alg_compare}
\end{figure}

{
Both the Layered-MADDPG and BO-MADDPG algorithms have a hierarchical structure that decomposes the network formation and trajectory optimization into two sub-problems. Here we also compare them with the conventional MADDPG algorithm that jointly adapts the UAVs' network formation and trajectories simultaneously, denoted as the Joint-MADDPG in Fig.~\ref{fig:alg_compare}. Each UAV's action ${\bf a}_i(t)$ in the Joint-MADDPG includes the flying direction ${\bf d}_i(t)$, the speed $v_i(t)$, and the network formation strategy $\phi_{i,j}(t)$ in each time step. The convergence performance of different MADDPG algorithms are shown in Fig.~\ref{fig:alg_compare}(a). It is clear that the BO-MADDPG converges faster and achieves a higher reward than the other algorithms. The reward of the BO-MADDPG sharply increases in the early learning stage, and the learning curve is more stable and smooth comparing to that of the Layered-MADDPG and the Joint-MADDPG algorithms. The Joint-MADDPG needs more iterations to obtain an effective strategy. In Fig.~\ref{fig:alg_compare}(b), we also evaluate the variance of the reward values during the learning process. A higher variance means that the learning algorithm can be unstable due to the random exploration. Fig.~\ref{fig:alg_compare}(b) shows that the BO-MADDPG has a comparable fluctuation as that of the Layered-MADDPG and the Joint-MADDPG algorithms in the early learning stage. However, the BO-MADDPG quickly converges to a stable value with a much smaller variance, while the Joint-MADDPG has a highly fluctuating reward performance, indicating the instability issues in training.}

\begin{figure}[t]
\centering
\subfloat[Adaptive network formation]{\includegraphics[width=0.45\linewidth]{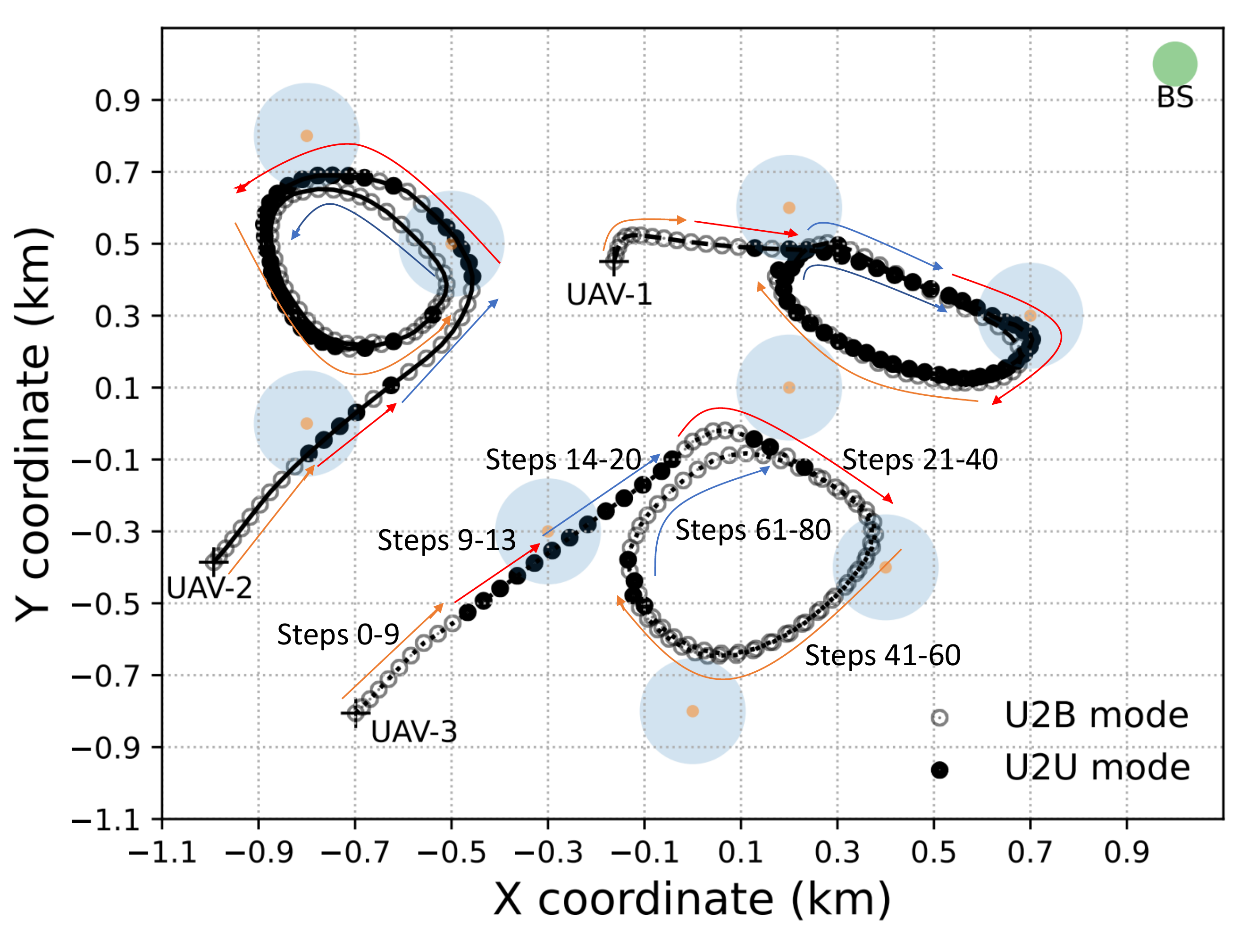}}
\subfloat[The UAVs' buffer sizes]{\includegraphics[width=0.45\linewidth]{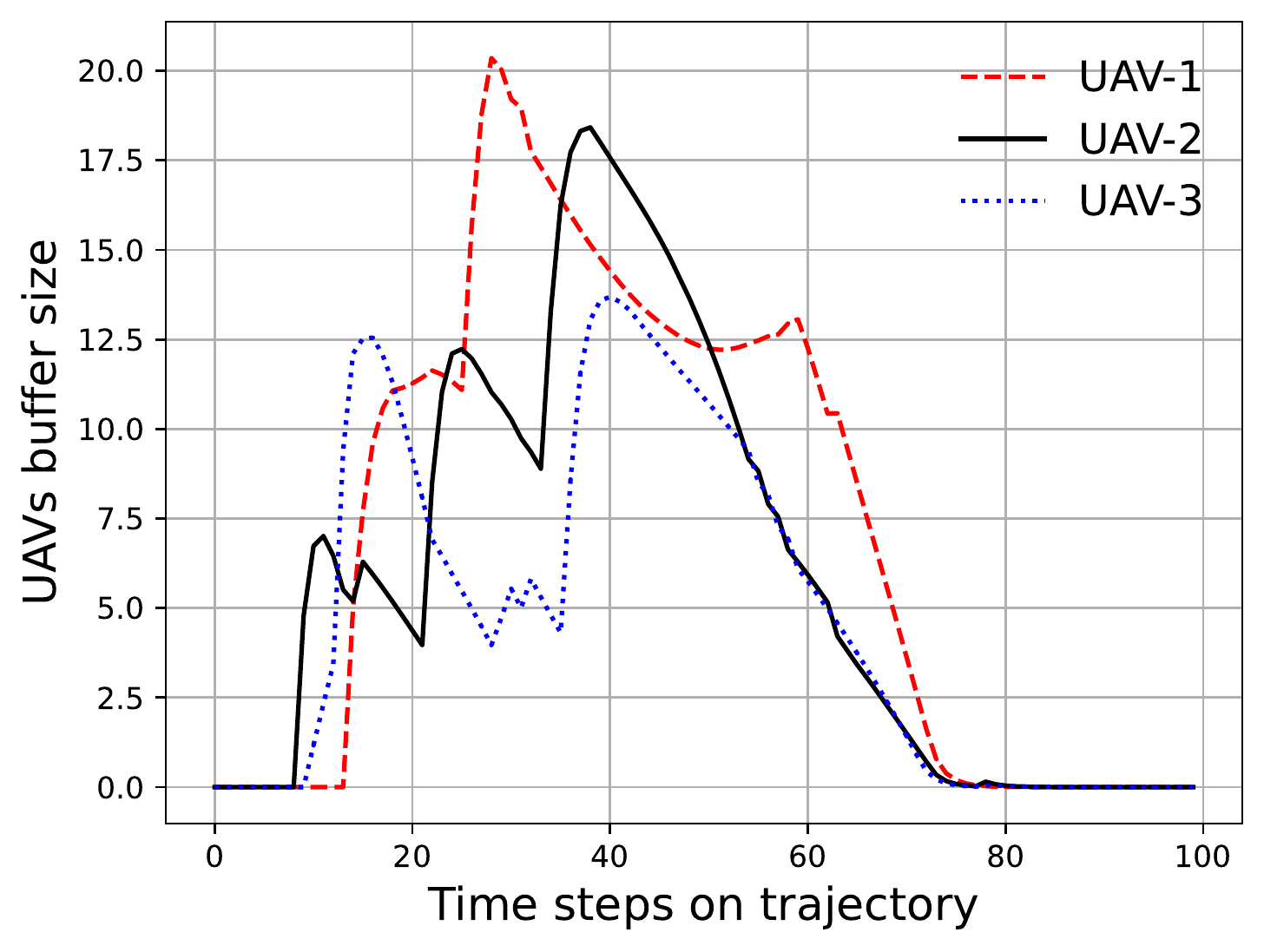}}
\caption{Adaptive network formation along the UAVs' trajectories.}
\label{fig:dy_NF}
\end{figure}

\subsection{Trajectory Planning with Adaptive Network Formation}

In this part, we evaluate the UAVs' trajectory planning with the energy-and-delay-aware adaptive network formation, i.e., Algorithm~\ref{alg-eda-algorithm}. Fig.~\ref{fig:dy_NF}(a) illustrates the trajectories of three UAVs and also reveals the dynamic change of the network formation as the UAVs fly along their trajectories. We use different types of marker points to indicate the U2U and U2B {communication modes}. The hollow circles represent the direct U2B communications. As the UAV moves away from the BS, the UAVs prefer to communicate with the next-hop UAVs via U2U communications. This requires a nearby UAV to act as the relay node and forward the data to the BS. As such, we can improve the data rate and reduce the transmission delay. As the UAVs fly closer to the BS, the U2B communication becomes more preferable, we can observe a switch from the U2U communication to the U2B communication. As shown in Fig.~\ref{fig:dy_NF}(a), the UAV-$1$'s service area is closer to the BS, while the UAV-$2$ and UAV-$3$'s service area is distant from the BS. Therefore, the UAV-$1$ will act as the relay node to assist the UAV-$2$'s or the UAV-$3$'s data offloading to the BS by using the high-speed U2U link between them. When the UAV-$3$ approaches the BS and its U2B channel becomes better, the UAV-$3$ disconnects the U2U link with the UAV-$1$. Instead, it chooses to offload its data through the U2B link directly. The UAV-$3$ can also act as a relay node for the UAV-$2$ when the UAV-$1$'s workload becomes excessive.
{ Specifically, in time steps 0-8 on the UAVs' trajectory, denoted as T0-8 in Fig.~\ref{fig:dy_NF}(a), the GUs are out of the UAVs' sensing range. In T9-13 of Fig.~\ref{fig:dy_NF}(a), the UAV-$2$ can collect sensing data from the GU-$2$, while the UAV-$3$ and the UAV-$1$ are still out of sensing range. As such, the UAV-$3$ can act as the relay node to assist the UAV-$2$'s data offloading through the U2U channel. In T14-20 of Fig.~\ref{fig:dy_NF}(a), the UAV-$2$ will disconnect its U2U link and switch to the direct U2B link, while the UAV-$3$ switches to the U2U link and selects the UAV-$1$ as its relay node. The UAV-$1$ becomes the closest to the BS and thus its sensing data can be quickly forwarded to the BS by using the U2B link. A similar analysis can apply to the remaining time steps in Fig.~\ref{fig:dy_NF}(a).}

In Fig.~\ref{fig:dy_NF}(b), we show the evolution of the UAVs' buffer sizes along the trajectories. Initially, the UAV-$1$'s buffer size is kept at a very low level as it is closer to the BS and has a more preferable U2B channel condition. The UAV-$1$'s sensing data can be quickly offloaded to the BS with the U2B channel. For the other two UAVs, they are distant from the BS and hence their buffer sizes increase significantly by using the low-rate U2B channels. When their buffer sizes continue to increase, the UAVs' cost functions become divergent. This drives the change of network formation by establishing the U2U links between the UAV-$2$/$3$ and the UAV-$1$. As shown in Fig.~\ref{fig:dy_NF}(b), we can observe the increase of the UAV-$1$'s buffer size while the corresponding decrease of the UAV-$2$'s and the UAV-$3$'s buffer sizes. Through the U2U communications, the UAV-$1$ will receive the data offloading from the other two UAVs. Finally, all UAVs complete the data offloading at the same time. This can minimize the total hovering time of all UAVs and stabilize their resource consumption. 

\begin{figure}[t]
\centering
\subfloat[The UAVs' decreasing costs]{\includegraphics[width=0.45\linewidth]{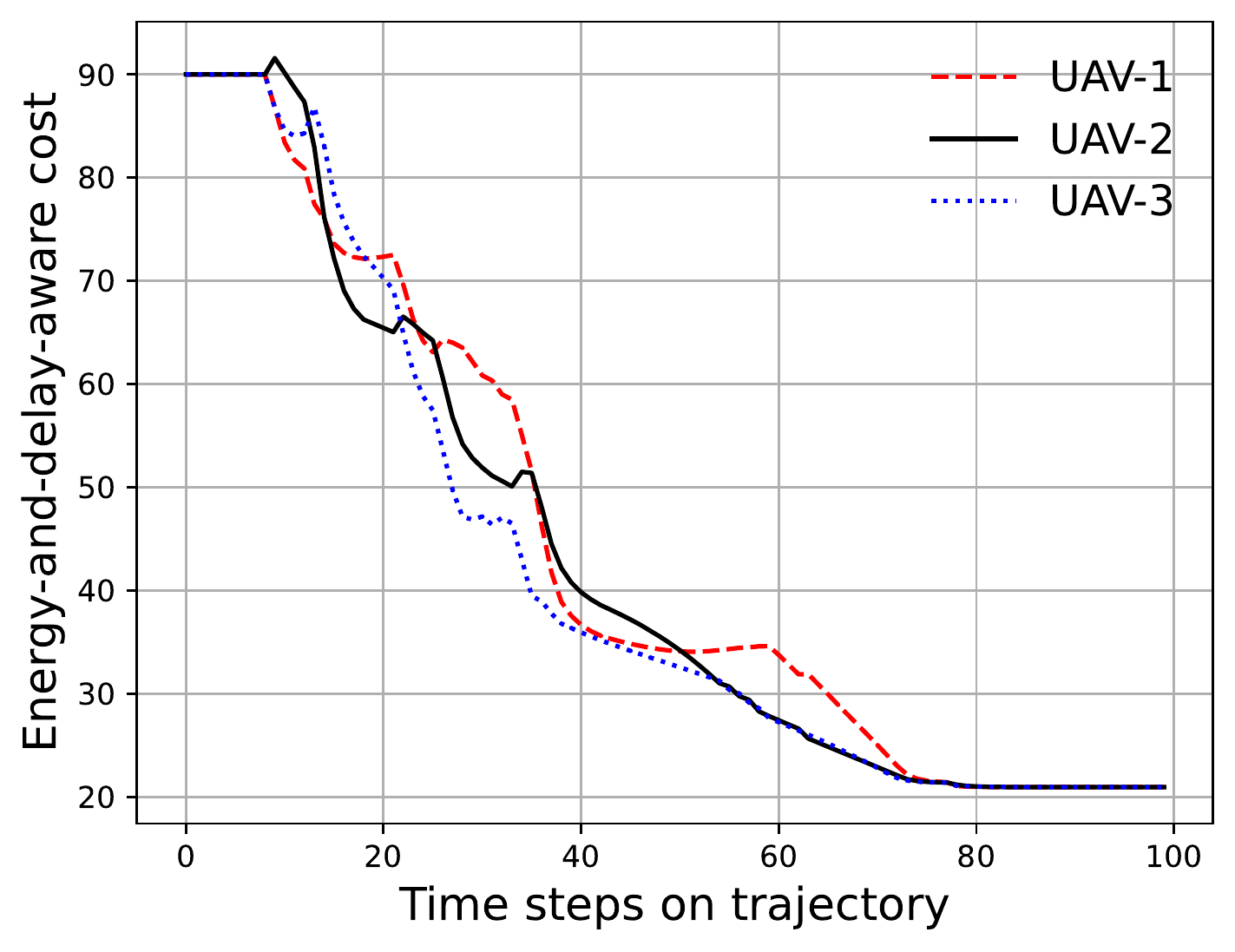}}
\subfloat[Load balance among UAVs]{\includegraphics[width=0.45\linewidth]{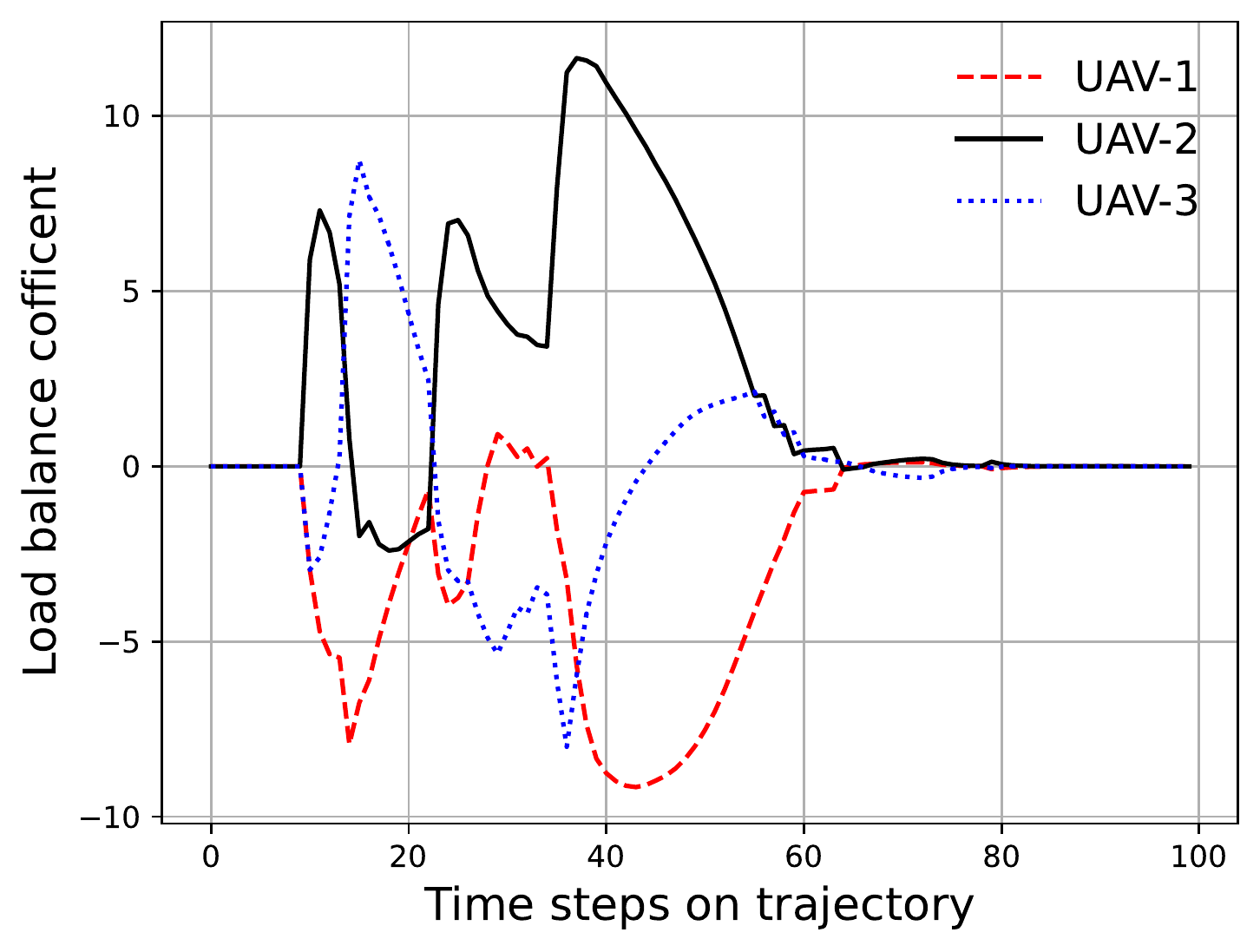}}
\caption{Reduced costs and enhanced load balance in Algorithm~\ref{alg-eda-algorithm}.}
\label{fig:data_time_change}
\end{figure}

In Fig.~\ref{fig:data_time_change}, we show the evolution of the UAVs' cost functions and the load balance coefficients that drive the UAVs' adaptive network formation in Algorithm~\ref{alg-eda-algorithm}. A larger cost value $c_i(t)$ implies that the UAV-$i$ demands more energy and incurs excessive transmission delay for data offloading, while the load balance coefficient $b_i(t)$ characterizes the direct U2B transmission capability. The UAVs' network formation aims to balance the UAVs' resource demands and transmission capabilities by equalizing the two coefficients $(c_i, b_i)$ of different UAVs, e.g.,~the UAV-$i$ with a larger $b_i(t)$ can establish the U2U link to a neighboring UAV with a lighter workload. As shown in Fig.~\ref{fig:data_time_change}(a), the UAVs' cost values decrease gradually by adapting the U2U links among different UAVs. When the UAV-$1$ acts as the relay node to assist the UAV-$2$'s data offloading, the UAV-$2$'s cost value drops rapidly while the UAV-$1$'s cost drops slowly as it receives extra workload from the UAV-$2$. When the UAV-$2$'s cost value becomes smaller than that of the UAV-$1$ up to some threshold, the UAV-$2$ will disconnect the U2U channel and switch back to the U2B channel to reduce the UAV-$1$'s workload. This process continues as all UAVs' cost values decrease to a small value when all GUs complete the data offloading tasks. Besides, the difference of the UAVs' cost values are maintained at a small range, which implies that the proposed adaptive network formation algorithm ensures a fair resource consumption among different UAVs. In Fig.~\ref{fig:data_time_change}(b), we show the evolution of the UAVs' load balance coefficients along the trajectories. As the UAV-$1$ is closer to the BS and has a higher transmission rate via the U2B link, initially it has a much smaller load balance coefficient $b_1$ than the other UAVs. The UAV-$2$'s workload is initially high but the load balance coefficient $b_2$ drops quickly as it offloads buffered data to the UAV-$1$ through the U2U channels.

\subsection{Comparison with Different Network Formation Strategy}
In this part, we verify that the adaptive network formation method in Algorithm~\ref{alg-eda-algorithm} can improve the reward performance of the UAVs' trajectory learning, comparing with several baseline network formation strategies. The first baseline is the non-cooperative strategy, which is commonly studied in the literature. It assumes that all UAVs have direct U2B connections to offload buffered data to the BS. The second baseline is similar to the multi-hop UAV network in~\cite{he2021multi}, which allows the distant UAV to offload its workload via multi-hop relay communications to the BS. To motivate the UAVs' relay communications, we devise a buffer-based algorithm that allows the UAVs to establish U2U connections with nearby UAVs when their buffer sizes exceed some threshold value. The data queue backlog sizes are the commonly used network information for routing or transmission control in multi-hop wireless networks~\cite{buffer-size}. The third baseline is the dynamic network formation algorithm (denoted as dynamic-NF) proposed in our previous work~\cite{9780862}. It maintains a cost value for each UAV based on its buffer size, remaining data, and energy consumption. Each UAV can adapt its U2U connections based on a comparison of its own cost value and the costs of the one-hop neighboring UAVs. Besides the cost value in~\cite{9780862}, the load balance coefficient is devised in Algorithm~\ref{alg-eda-algorithm} to characterize each UAVs' transmission capability.


\begin{figure}[t]
\centering
\subfloat[Reward dynamics in learning]{\includegraphics[width=0.45\linewidth]{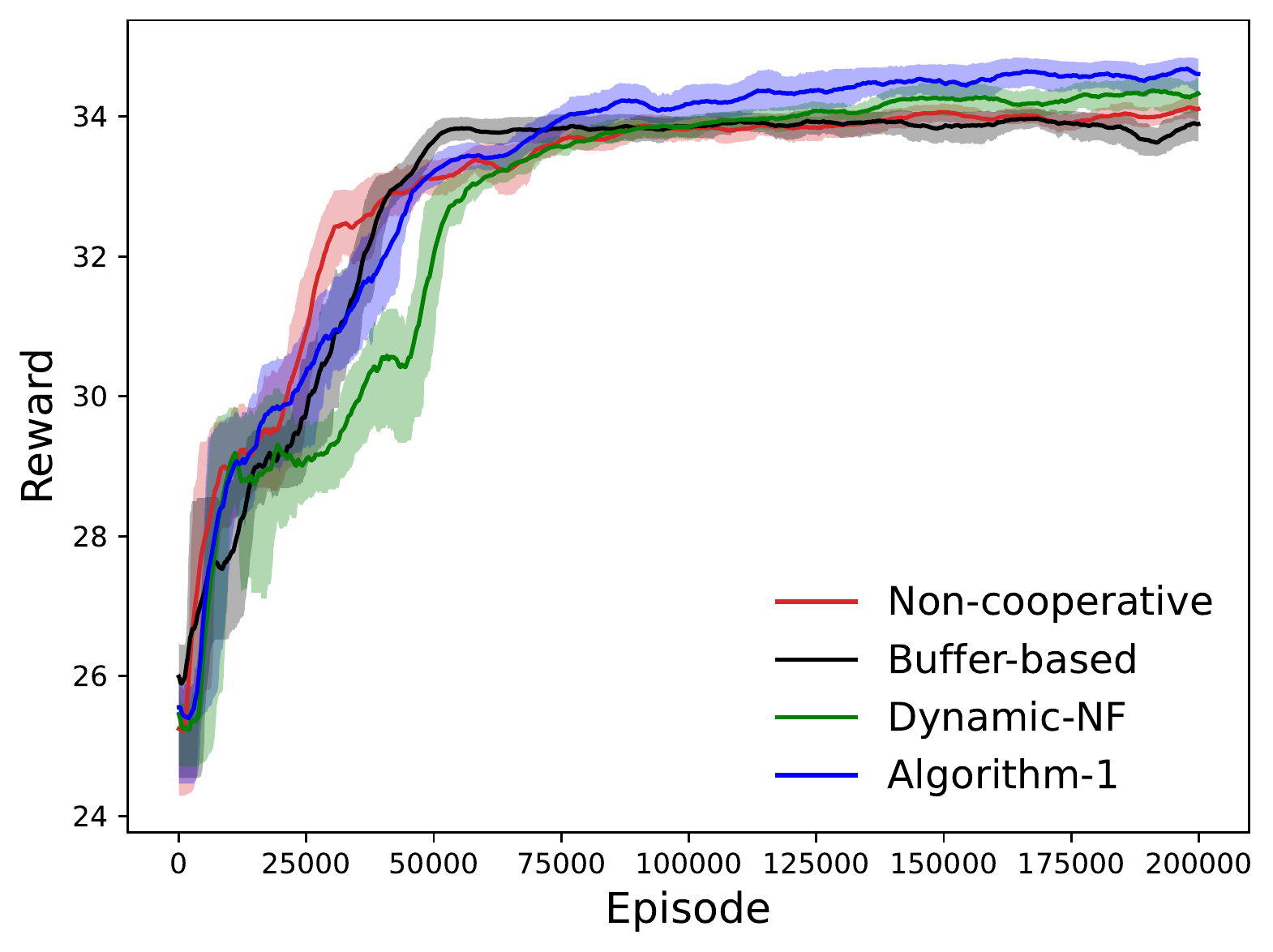}}
\subfloat[UAVs' buffer sizes]{\includegraphics[width=0.45\linewidth]{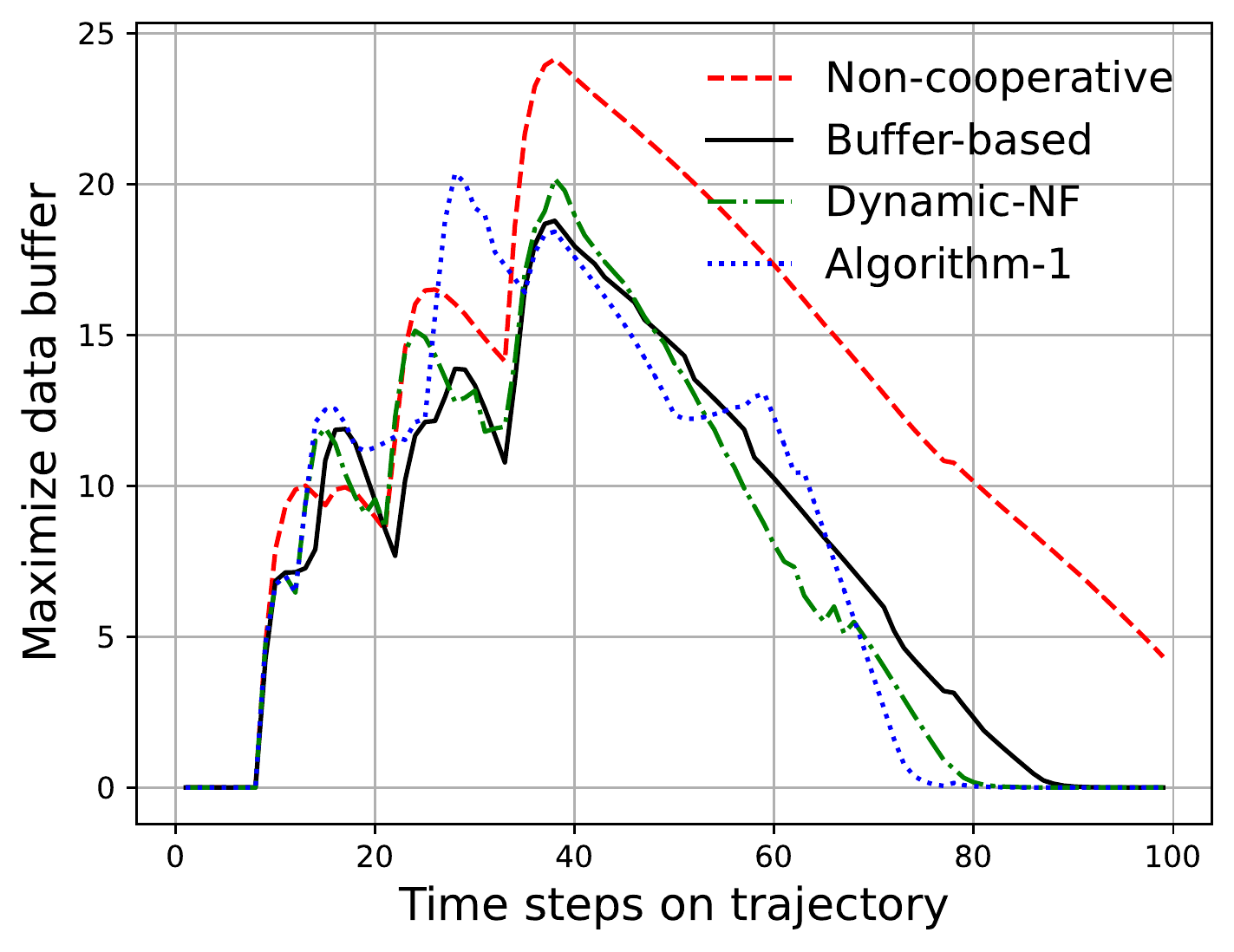}}
\caption{Adaptive network formation improves the reward and offloading performance for trajectory learning.}
\label{fig:algorithm-compare1}
\end{figure}

In Fig.~\ref{fig:algorithm-compare1}(a), we compare the convergence and reward performance of different network formation algorithms in the UAVs' trajectory planning. The non-cooperative strategy omits the UAVs' task cooperation and shows the fastest convergence speed, but with the cost of a lower reward performance in trajectory learning. The buffer-based adaptive network formation strategy tries to exploit the UAVs' task cooperation. It omits other critic resource limits (e.g.,~the UAVs' energy status and the GUs' workload demands) that may affect the UAVs' trajectories. Comparing to the non-cooperative strategy and the buffer-based algorithm, the Dynamic-NF algorithm achieves a higher reward but with a slower convergence rate. Our Algorithm~\ref{alg-eda-algorithm} achieves both better reward and faster convergence than that of the Dynamic-NF algorithm as shown in Fig.~\ref{fig:algorithm-compare1}(a). In Fig.~\ref{fig:algorithm-compare1}(b), we show the UAVs' maximum buffer size along the UAVs' trajectories. The UAVs' data buffers can be maintained at a low level by adapting the UAVs' network formation, whereas the non-cooperative strategy results in the largest buffer size, especially for a distant UAV with the worst U2B channel conditions to the BS. The buffer-based algorithm merely focuses on the UAVs' buffer size. Hence, it can maintain the lowest buffer size as shown in Fig.~\ref{fig:algorithm-compare1}(b). Algorithm~\ref{alg-eda-algorithm} leads to a comparable maximum buffer size with the buffer-based algorithm. However, it can complete all UAVs' data offloading in a shorter {time comparing} with the other baselines. This implies that Algorithm~\ref{alg-eda-algorithm} can reduce the UAVs' hovering time and energy consumption in the air.

\begin{figure}[t]
\centering
\subfloat[Remaining data in the system]{\includegraphics[width=0.45\linewidth]{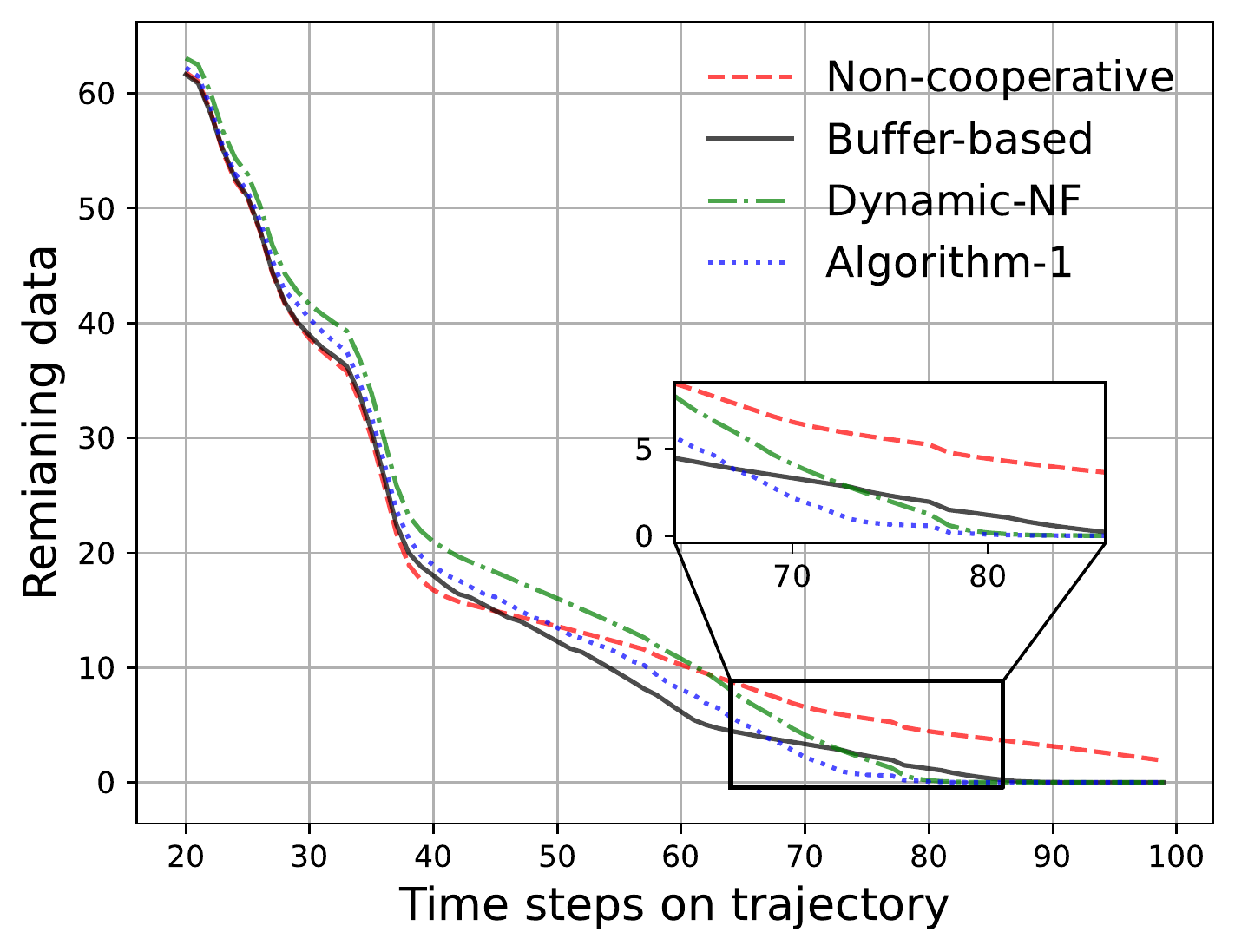}}
\subfloat[Workload completion time]{\includegraphics[width=0.45\linewidth]{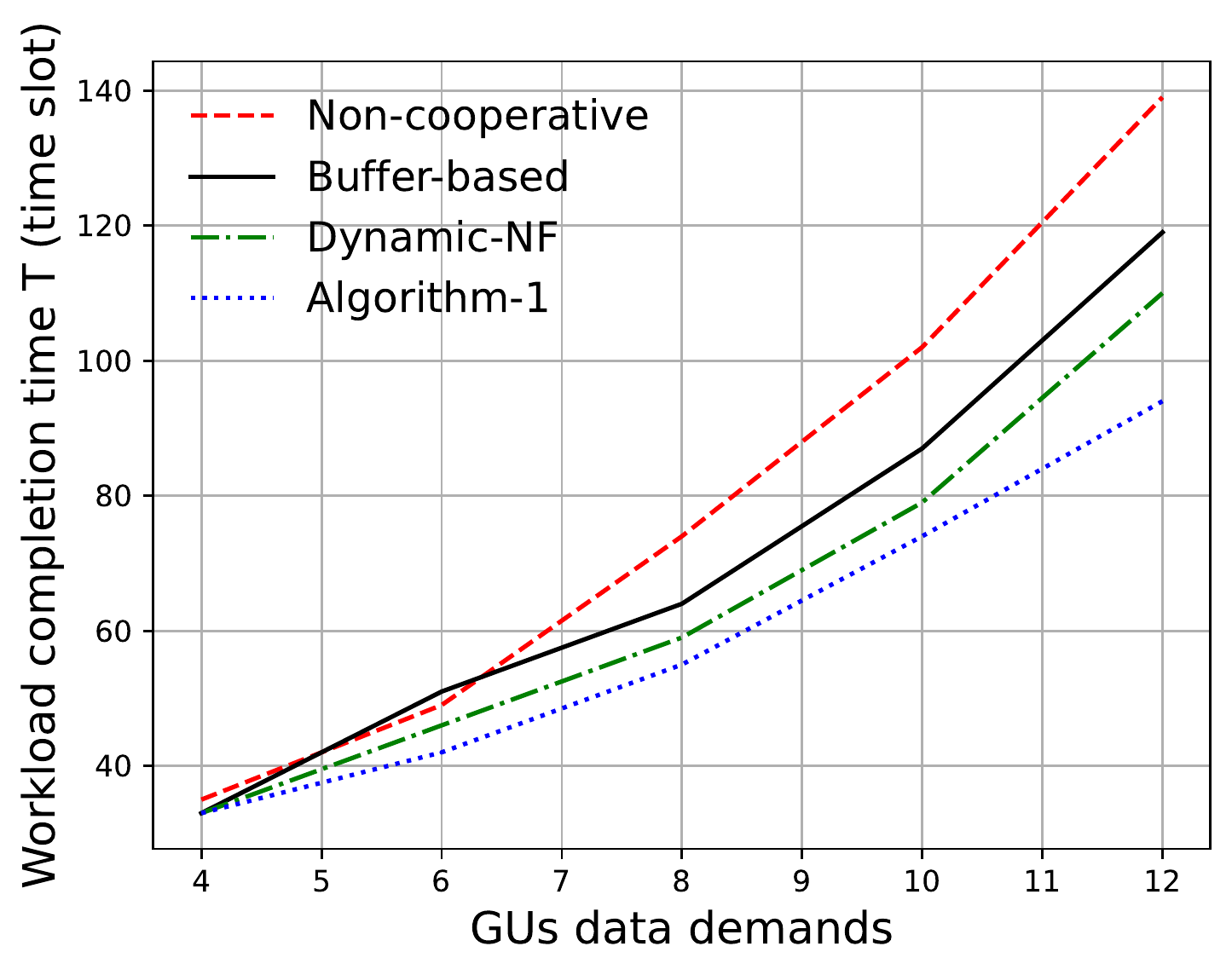}}
\caption{Adaptive network formation improves the time efficiency for the GUs' data offloading and balances the UAVs' workload.}
\label{fig:algorithm-compare}
\end{figure}

In Fig.~\ref{fig:algorithm-compare}(a), we show the overall data remaining in the GUs' and the UAVs' buffers. The common observation is that the remaining data in the system gradually decreases in all algorithms. An interesting observation is that the non-cooperative strategy has a faster decreasing rate initially, but the rate of decreasing slows down and shows a {fat tail}. This implies a longer workload completion time due to the rate-limited U2B channels. The buffer-based and dynamic-NF algorithms can improve the system capacity by using the high-speed U2U links. The workload completion time can be significantly reduced comparing to the non-cooperative strategy. Algorithm~\ref{alg-eda-algorithm} achieves the best offloading performance as shown in Fig.~\ref{fig:algorithm-compare}(a). With low workload demands from the GUs, Algorithm~\ref{alg-eda-algorithm} achieves a similar workload completion time as that of the buffer-based and dynamic-NF algorithms. We expect that Algorithm~\ref{alg-eda-algorithm} can perform much better as the GUs' workload demands increase. In Fig.~\ref{fig:algorithm-compare}(b), we compare the workload completion time of different network formation strategies as we gradually increase the GUs' workload demands. It is clear that Algorithm~\ref{alg-eda-algorithm} can complete transmission task in a shorter time and the performance gain becomes more significant compared with the baselines. This implies that our Algorithm~\ref{alg-eda-algorithm} can be more robust to balance the UAVs' workload and resource demands by dynamically {adapting} the network formation along the UAVs' trajectories.

\section{Conclusions and Future Works}\label{sec-conclusion}

In this paper, we have proposed a {layered learning} algorithm to jointly optimize the UAVs' network formation and trajectories for wireless data offloading. Given the UAVs' trajectories, the network formation is adapted mainly based on the UAVs' energy consumption and buffer sizes. When the network formation is changed, we further update the UAVs' trajectories by using the {multi-agent learning} algorithm. Moreover, to improve the learning efficiency, we employ Bayesian optimization to estimate the UAVs' flying decisions based on historical information. This helps avoid ineffective action explorations and improve the convergence in learning. Simulation results have demonstrated that our algorithm can adapt the UAVs' network formation along their trajectories and therefore collect the sensing data more efficiently.

\bibliographystyle{IEEEtran}

\bibliography{reference}

\end{document}